\documentclass[letterpaper]{article} % DO NOT CHANGE THIS
\usepackage{aaai2026}  % DO NOT CHANGE THIS
\usepackage{times}  % DO NOT CHANGE THIS
\usepackage{helvet}  % DO NOT CHANGE THIS
\usepackage{courier}  % DO NOT CHANGE THIS
\usepackage[hyphens]{url}  % DO NOT CHANGE THIS
\usepackage{graphicx} % DO NOT CHANGE THIS
\usepackage{natbib}  % DO NOT CHANGE THIS AND DO NOT ADD ANY OPTIONS TO IT
\usepackage{caption} % DO NOT CHANGE THIS AND DO NOT ADD ANY OPTIONS TO IT
\usepackage{algorithm}
\usepackage{algorithmic}
\usepackage{adjustbox}
\usepackage{longtable}
\usepackage{array} 
\usepackage{booktabs}
\usepackage{multirow}
\usepackage{xcolor}
\usepackage{colortbl}
\usepackage{amsfonts}
\usepackage[most]{tcolorbox}
\usepackage{amssymb}

\newcommand{\greencheck}{\textcolor{green!70!black}{\checkmark}}
\newcommand{\redx}{\textcolor{red!70!black}{\texttimes}}% for \checkmark
\usepackage{newfloat}
\usepackage{listings}
\DeclareCaptionStyle{ruled}{labelfont=normalfont,labelsep=colon,strut=off} % DO NOT CHANGE THIS
\floatstyle{ruled}
\newfloat{listing}{tb}{lst}{}
\floatname{listing}{Listing}
\nocopyright
\title{MSU-Bench: Towards Understanding the Conversational Multi-talker Scenarios}
\author{Shuai Wang\textsuperscript{\rm 1}\equalcontrib, Zhaokai Sun\textsuperscript{\rm 2}\equalcontrib, Zhennan Lin\textsuperscript{\rm 2}\equalcontrib, Chengyou Wang\textsuperscript{\rm 2}\equalcontrib, Zhou Pan\textsuperscript{\rm 3}, Lei Xie\textsuperscript{\rm 2}\thanks{Corresponding author}}
\affiliations{
    \textsuperscript{\rm 1}School of Intelligence Science and Technology, Nanjing University
    
    \textsuperscript{\rm 2}Audio, Speech and Language Processing Lab (ASLP@NPU), Northwestern Polytechnical University
    
    \textsuperscript{\rm 3}Li Auto Inc. \\
    shuaiwang@nju.edu.cn, \{zksun,znlin,asd6404112a\}@mail.nwpu.edu.cn, lxie@nwpu.edu.cn
}

\begin{document}

\maketitle

\begin{abstract}

Spoken Language Understanding (SLU) has progressed from traditional single-task methods to large audio language model (LALM) solutions. Yet, most existing speech benchmarks focus on single-speaker or isolated tasks, overlooking the challenges posed by multi-speaker conversations that are common in real-world scenarios. We introduce \textbf{MSU-Bench}, a comprehensive benchmark for evaluating multi-speaker conversational understanding with a speaker-centric design. Our hierarchical framework covers four progressive tiers: single-speaker static attribute understanding, single-speaker dynamic attribute understanding, multi-speaker background understanding, and multi-speaker interaction understanding. This structure ensures all tasks are grounded in speaker-centric contexts, from basic perception to complex reasoning across multiple speakers.
By evaluating state-of-the-art models on MSU-Bench, we demonstrate that as task complexity increases across the benchmark's tiers, all models exhibit a significant performance decline. We also observe a persistent capability gap between open-source models and closed-source commercial ones, particularly in multi-speaker interaction reasoning. These findings validate the effectiveness of MSU-Bench for assessing and advancing conversational understanding in realistic multi-speaker environments. Demos can be found in the supplementary material.

\end{abstract}

\begin{links}
    \link{Repo}{https://github.com/ASLP-lab/MSU-Bench}
    \link{Demo}{https://aslp-lab.github.io/msu-bench.github.io/}
    % \link{Extended version}{https://aaai.org/example/extended-version}
\end{links}

Spoken Language Understanding (SLU) constitutes a fundamental task in artificial intelligence, enabling machines to interpret human speech beyond mere transcription. Recent advances in SLU research have transitioned from traditional single-task approaches, such as automatic speech recognition (ASR), automatic speaker verification (ASV), and spoken sentiment analysis (SSA), toward Large Audio Language Models (LALMs)~\cite{peng2024survey,su2025audiosurvey}. Building upon established Large Language Model (LLM) paradigms, sophisticated LALMs such as  LTU-AS~\cite{gong2023joint}, Salmonn~\cite{tang2024salmonn}, Qwen-Audio~\cite{chu2023qwen,chu2024qwen2audio}, and OSUM~\cite{geng2025osum} have emerged, demonstrating exceptional general speech understanding capabilities. 

However, real-world conversations inherently involve multiple speakers and present fundamentally different challenges than single-speaker scenarios. Human dialogues are inherently collaborative and social, involving complex interactions among multiple participants where speakers frequently interrupt one another, reference previous statements, and dynamically shift conversational roles. While researchers have developed sophisticated techniques for individual aspects of multi-speaker processing, including speaker diarization, speech separation, and target speaker extraction, these methods predominantly focus on isolated technical problems without capturing the holistic complexity of conversational dynamics. Critical speaker-centric phenomena such as social role analysis, power dynamics, and interactional patterns remain largely unexplored.

Existing speech benchmarks~\cite{chen2024voicebench,yang2024air,ao2024sd,sakshimmau,wang2025mmsu,wang2025audiobench} typically aggregate speaker-related tasks with general speech or dialogue tasks, rarely isolating the unique challenges inherent in speaker-centric understanding within authentic conversational contexts. Consequently, essential multi-speaker dynamics, including dominance detection, turn-taking analysis, and social role identification, remain underexplored. This gap is particularly significant given that most real-world applications require understanding not just individual speakers but the complex interplay between multiple participants in dynamic conversational settings.
\begin{table*}[!htbp]
\centering

\renewcommand{\arraystretch}{1.4}
\setlength{\tabcolsep}{6pt}
\small
\begin{tabular}{@{}lccccccc@{}}
\toprule
\multirow{2}{*}{\textbf{Characteristics}} & \multicolumn{7}{c}{\textbf{Speech Understanding Benchmarks}} \\
\cmidrule(l){2-8}
 & \textbf{VoiceBench} & \textbf{MMSU} & \textbf{MMAU} & \textbf{AudioBench} & \textbf{AIR-Bench} & \textbf{SD-Eval} & \textbf{MSU-Bench} \\
\midrule
\rowcolor{gray!15}
\textbf{Speaker-Oriented}& \redx & \redx & \redx & \redx & \redx & \redx & \greencheck \\
\textbf{Audio Source}  & TTS+RPC & RPC & RPC & RPC & RPC & TTS+RPC & RPC \\
\rowcolor{gray!15}
\textbf{Conversation Type} & Mono. & Dial. & Dial. & Dial. & Dial. & Dial. & Dial. \\
\textbf{Multi-speaker} & \redx & \redx & \redx & \redx & \redx & \redx & \greencheck \\
\rowcolor{gray!15}
\textbf{Speaker-related Task} & \greencheck & \greencheck & \greencheck & \greencheck & \greencheck & \greencheck & \greencheck \\
\bottomrule
\end{tabular}
\caption{Comparative analysis of construction characteristics across seven speech understanding benchmarks. \greencheck~indicates support, \redx~indicates no support. TTS: Text-to-Speech, RPC: Real Person Recordings, Mono.: Monologue, Dial.: Dialogue.}
\label{tab:benchmark-comparison}
\vspace{-4mm}
\end{table*}

To bridge this critical gap, we introduce \textbf{MSU-Bench}, the first comprehensive benchmark specifically designed to define and evaluate multi-speaker understanding in authentic interactive scenarios. MSU-Bench employs a hierarchical framework that decomposes speaker-centric understanding into four progressive tiers of complexity: single-speaker static attribute understanding (e.g., speaker counting, demographic profiling), single-speaker dynamic attribute understanding (e.g., emotion state tracking, voice quality evolution), multi-speaker background understanding (e.g., venue/event inference, role identification), and multi-speaker interaction understanding (e.g., dominance detection, interruption pattern analysis). This systematic decomposition enables targeted evaluation of the social and interactive dimensions of conversational understanding while maintaining clear progression from basic perceptual tasks to complex reasoning scenarios.

Our work makes the following key contributions:
\begin{itemize}
    \item We present the first benchmark that is dedicated to conversational speaker-centric understanding. A four-tier hierarchical structure from basic perception to advanced reasoning is designed.
    \item We propose a comprehensive ``5M'' design. This covers multi-tier, multi-speaker, multi-lingual, multi-scenario, and multi-task aspects for thorough evaluation.
    \item We conduct extensive empirical analysis of both open-source and closed-source models. Our results show persistent performance gaps, especially in multi-speaker interaction reasoning.
    \item We provide a detailed task construction pipeline and open-source codebase. This supports reproducibility and promotes future research on speaker-centric understanding in conversations.
\end{itemize}

\section{Related Works}

\subsection{Speaker Modeling and Analysis}
Speaker information constitutes a crucial dimension of the acoustic signal~\cite{wang2024overview}. Conventionally, the scope of speaker modeling has been narrowly centered on identity for tasks like recognition and verification. However, a more holistic perspective, often termed speaker understanding, extends to a rich set of paralinguistic traits, including a speaker's accent, age, and emotional state. While early research addressed these characteristics through separate, task-specific systems, a recent paradigm shift has led to the development of benchmarks that evaluate speaker attributes more comprehensively. For instance, the VoxProfile~\cite{feng2025vox} benchmark was introduced to analyze speaker profiles across various facets. A limitation of such approaches is their inherent focus on single-utterance, single-speaker scenarios, which precludes the analysis of more complex, interactive contexts that are essential for understanding real-world conversational dynamics.
\subsection{Speech Understanding Models}
Speech understanding encompasses the machine interpretation of semantic content and emotional nuances in spoken language, extending beyond simple transcription to include intent recognition, sentiment analysis, and dialogue act identification. Recent advances in large language models have significantly enhanced this capability by integrating audio modalities with LLMs~\cite{huang2024audiogpt,zhang2023speechgpt,ghosh2025audio,goel2025audio,chu2024qwen2audio,xu2025qwen2}. Current approaches fall into two categories: cascade and end-to-end methods. Cascade approaches utilize automatic speech recognition followed by natural language processing, as exemplified by AudioGPT~\cite{huang2024audiogpt}, combining Whisper with LLMs. While modular and industrially mature, this method suffers from error propagation and acoustic information loss. End-to-end approaches directly map speech signals to semantic representations, demonstrated by models such as SpeechGPT~\cite{zhang2023speechgpt}, Salmonn~\cite{DBLP:conf/iclr/TangYSC000M024,DBLP:journals/corr/abs-2505-17060}, Glm-4-voice~\cite{zeng2024glm}, GPT
4o-Audio, Gemini~\cite{team2023gemini}, Kimi-Audio~\cite{ding2025kimi}, Step-audio~\cite{wu2025step,huang2025step} and the Qwen-Audio series~\cite{chu2023qwen,chu2024qwen2}. These models exhibit greater robustness and universal audio understanding capabilities. Recent multimodal extensions like Gemini and Qwen2.5-Omni further integrate audio and visual information.

\subsection{Speech Understanding Benchmarks} 

Recent advances in LALMs have catalyzed the emergence of diverse benchmarks for speech understanding. To systematically assess the current landscape, we compare several representative benchmarks with our proposed MSU-Bench in Table~\ref{tab:benchmark-comparison}. Early efforts such as  AudioBench~\cite{wang2025audiobench} mainly target foundational capabilities, including automatic speech recognition (ASR) and audio classification. More recent benchmarks, such as MMAU~\cite{sakshimmau} and MMSU~\cite{wang2025mmsu}, extend the scope to encompass audio-based question answering and the assessment of paralinguistic features. Nevertheless, as summarized in Table~\ref{tab:benchmark-comparison}, none of the existing benchmarks are explicitly speaker-centric or designed to evaluate multi-speaker interactions. While many adopt dialogue recordings (Dial.) and include speaker-aware tasks, they fail to address the critical challenge of modeling the relationships and interaction logic among different speakers, which is essential for a comprehensive conversational understanding. To bridge this gap, we introduce \textbf{MSU-Bench}, the first benchmark specifically dedicated to the rigorous evaluation of multi-speaker understanding in realistic conversational scenarios.

\section{MSU-Bench: Hierarchical Design for Multi-Speaker Understanding}

\begin{figure*}[!htb]
\centering
\includegraphics[width=0.78\textwidth]{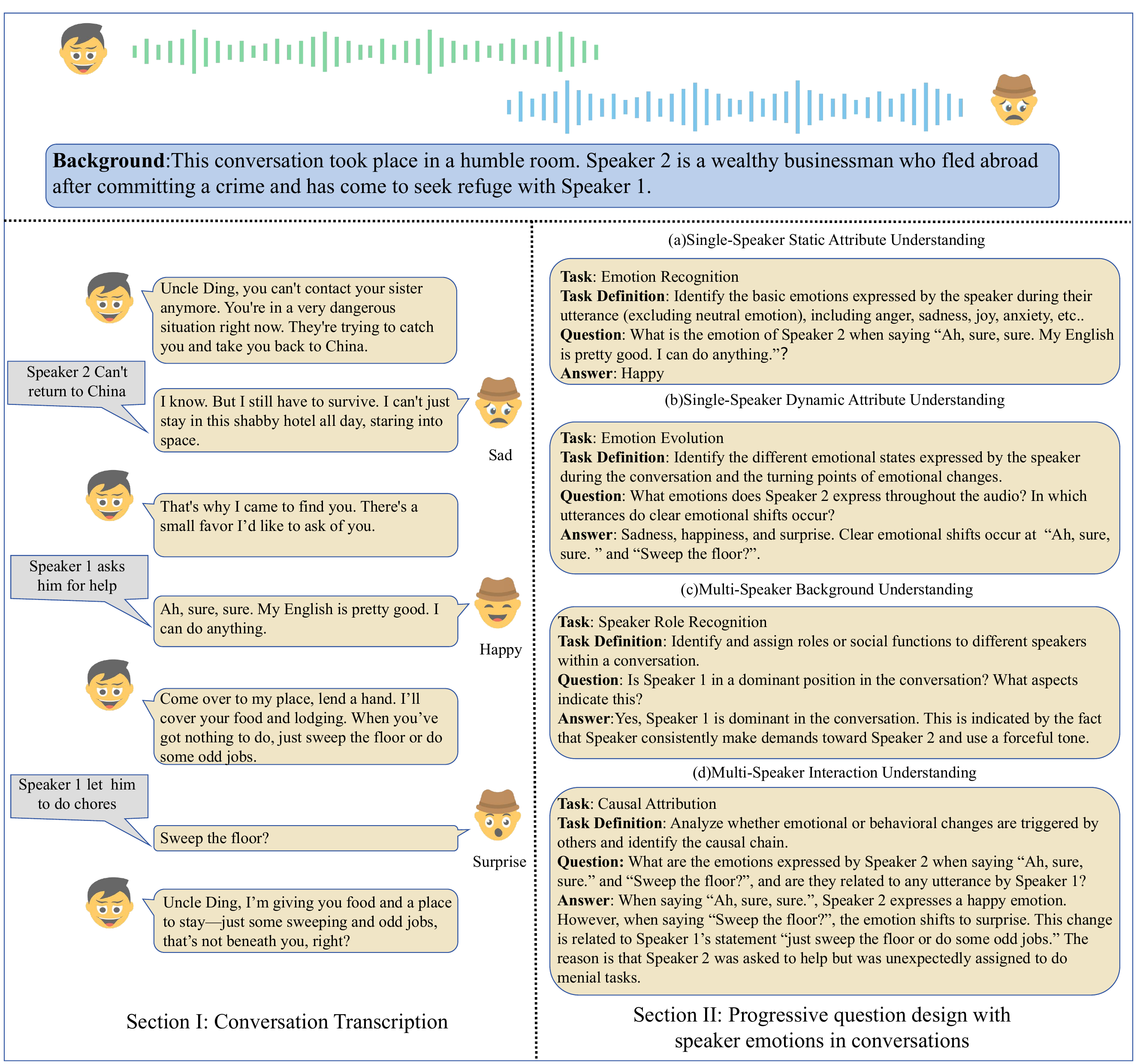}
\caption{\textbf{Example of MSU-Bench QA.} The conversation scenario involves two speakers with different emotional states and social dynamics, demonstrating the progression from basic emotion recognition to complex causal reasoning.}
\label{fig:example}
\vspace{-5mm}
\end{figure*}

\begin{table*}[!t]
\centering
\renewcommand{\arraystretch}{1.18}
\begin{tabular}{@{}p{0.024\textwidth} p{0.21\textwidth} p{0.39\textwidth} p{0.30\textwidth}@{}}
\toprule
& \textbf{Capability} & \textbf{Description} & \textbf{Representative Tasks} \\
\midrule
\multirow{30}{*}{\rotatebox{270}{\parbox{14cm}{\centering\footnotesize
\hspace{0.6cm}
\textbf{Single-Speaker $\rightarrow$ Multi-Speaker} \hspace{1.2cm}
\textbf{Static $\rightarrow$ Dynamic} \hspace{1.2cm}
\textbf{Perception $\rightarrow$ Reasoning}}}}
& \multicolumn{3}{c}{\cellcolor{gray!13}\textbf{Tier 1: Single-Speaker Static Attribute Understanding}} \\
\cmidrule(lr){2-4}
& \textbf{Speaker Recognition (SR)} &
Identify and track speakers in multi-speaker environments, focusing on speech content, timing, alternation, frequency, and interaction structure.
& \begin{tabular}[t]{@{}l@{}}
Speaker Recognition \\
Speaker Counting \\
Silence/Overlap Detection
\end{tabular} \\
\addlinespace
& \textbf{Speaker Attribute Comprehension (SAC)} &
Determine static attributes such as gender, age, accent, and language background of the speaker.
& \begin{tabular}[t]{@{}l@{}}
Accent/Dialect Recognition \\
Language Recognition \\
Gender Recognition \\
Age Recognition
\end{tabular} \\
\addlinespace
& \textbf{Speaker Paralinguistic Analysis (SPA)} &
Analyze vocal characteristics such as timbre, fluency, and emotional tone.
& \begin{tabular}[t]{@{}l@{}}
Voice Quality Analysis \\
Speech Flow Analysis \\
Emotion Recognition
\end{tabular} \\
\cmidrule(lr){2-4}
& \multicolumn{3}{c}{\cellcolor{gray!13}\textbf{Tier 2: Single-Speaker Dynamic Attribute Understanding}} \\
\cmidrule(lr){2-4}
& \textbf{Speaker Dynamic Analysis (SDA)} &
Detect and interpret dynamic changes in emotion, voice quality, and perspective over the course of a conversation.
& \begin{tabular}[t]{@{}l@{}}
Emotion Evolution \\
Voice Quality Evolution \\
Opinion Change Recognition
\end{tabular} \\
\addlinespace
& \textbf{Speaker Cultural Identity Integration (SCII)} &
Infer cultural background, geographical affiliation, age group, and cognitive style by analyzing language, accent, and expression preferences.
& \begin{tabular}[t]{@{}l@{}}
Language/Accent Cultural Reasoning \\
Expression Preference Recognition \\
Geographical Location Estimation
\end{tabular} \\
\cmidrule(lr){2-4}
& \multicolumn{3}{c}{\cellcolor{gray!13}\textbf{Tier 3:  Multi-Speaker Background Understanding}} \\
\cmidrule(lr){2-4}
\addlinespace
& \textbf{Multi-Speaker Scene Inference (MSSI)} &
Infer conversational venue and predict conversational outcomes from topics and language styles.
& \begin{tabular}[t]{@{}l@{}}
Dialogue Background Reasoning \\
Dialogue Result Reasoning \\
\end{tabular} \\
\addlinespace
& \textbf{Multi-Speaker Relationship Inference (MSRI)} &
Understand speaker relationships and infer social roles in multi-speaker conversations.
& \begin{tabular}[t]{@{}l@{}}
Speaker Role Recognition \\
Social Role Recognition
\end{tabular} \\
\cmidrule(lr){2-4}
& \multicolumn{3}{c}{\cellcolor{gray!13}\textbf{Tier 4:  Multi-Speaker Interaction Understanding}} \\
\cmidrule(lr){2-4}
& \textbf{Multi-Speaker Transcription (MST)} &
Identify and distinguish multiple speakers in conversations, ensuring accurate restoration of semantic content.
& Dialogue Transcription \\
\addlinespace
& \textbf{Multi-Speaker Interaction Analysis (MSIA)} &
Understand interpersonal dynamics in multi-speaker interactions through paralinguistic and social cues.
& \begin{tabular}[t]{@{}l@{}}
Paralinguistic Interaction Analysis \\
Social Interaction Analysis
\end{tabular} \\
\addlinespace
& \textbf{Multi-Speaker Contextual Reasoning (MSCR)} &
Analyze emotional shifts, intention changes, and interaction logic among speakers, enabling semantic-based cross-speaker reasoning.
& \begin{tabular}[t]{@{}l@{}}
Causal Attribution \\
Motivation Reasoning
\end{tabular} \\
\bottomrule
\end{tabular}
\caption{
\textbf{Hierarchical Multi-Speaker Understanding Tasks.}
The framework systematically progresses from single-speaker static perception to multi-speaker dynamic reasoning, encompassing 10 core capabilities and 25 representative tasks. Detailed descriptions and examples of individual tasks can be found in the appendix.
}
\label{tab:tiersep}
\end{table*}

\begin{table*}[ht]
\centering
\renewcommand{\arraystretch}{1.15}
\setlength{\tabcolsep}{5.0pt}
\rowcolors{2}{gray!10}{white}
\small 
\begin{tabular}{l|cccc|ccc|ccc|cccc|c}
\toprule
\multirow{2}{*}{\textbf{Models}} & \multicolumn{4}{c|}{\textbf{Tier 1}} & \multicolumn{3}{c|}{\textbf{Tier 2}} & \multicolumn{3}{c|}{\textbf{Tier 3}} & \multicolumn{4}{c|}{\textbf{Tier 4}} & \multirow{2}{*}{\textbf{Avg}} \\
\cmidrule(lr){2-5} \cmidrule(lr){6-8} \cmidrule(lr){9-11} \cmidrule(lr){12-15}
 & SR & SAC & SPA & Avg & SDA & SCII & Avg & MSSI & MSRI & Avg & MST & MSIA & MSCR & Avg & \\
\midrule
Kimi-Audio        & 0.39 & 0.53 & 0.38 & 0.44 & 0.21 & 0.29 & 0.25 & 0.38 & 0.40 & 0.39 & 0.35 & 0.23 & 0.24 & 0.25 & 0.35 \\
Qwen2.5-Omni      & 0.48 & 0.48 & 0.37 & 0.45 & 0.26 & 0.34 & 0.29& 0.33 & 0.44 & 0.36 & 0.29 & 0.34 & 0.26 & 0.30 & 0.37 \\
GPT-4o-Audio      & 0.52 & 0.65 & \textbf{0.55} & 0.58 & 0.38 & 0.52 & 0.44 & 0.70 & 0.51 & 0.64 & 0.37 & 0.49 & 0.36 & 0.43 & 0.52 \\
Gemini-2.5-Flash  & 0.49 & \textbf{0.70} & 0.51 & 0.58 & 0.41 & 0.57 & 0.48 & 0.76 & 0.66 & 0.73 & 0.38 & 0.51 & 0.39 & 0.45 & 0.55 \\
\rowcolor{gray!20}
Gemini-2.5-Pro    & \textbf{0.55} & \textbf{0.70} & 0.54 & \textbf{0.61} & \textbf{0.46} & \textbf{0.61} & \textbf{0.53} & \textbf{0.80} & \textbf{0.67} & \textbf{0.76} & \textbf{0.44} & \textbf{0.56} & \textbf{0.47} & \textbf{0.51} & \textbf{0.59} \\

\bottomrule
\end{tabular}

\caption{Performance comparison of different models across four tiers and various capabilities. Bold values indicate the best performance in each group. Definitions of all capability abbreviations are provided in Table 2. Note that the Avg values are computed on all involved cases, instead of simply averaging individual capability values.}
\label{tab:overall_results}
\vspace{-4mm}
\end{table*}

\subsection{Hierarchical Task Framework}

We propose a four-tier hierarchical framework for multi-speaker understanding tasks, organized by increasing complexity to systematically evaluate model capabilities across different tiers of conversational understanding. The definitions and scope of each tier are demonstrated in Table~\ref{tab:tiersep}.

\paragraph{Framework Overview.}
Our framework progresses from basic speaker-level perception to complex multi-party interaction reasoning, as Figure~\ref{fig:example}. The progression follows a natural cognitive hierarchy: Tier 1 establishes foundational recognition capabilities for static speaker attributes, Tier 2 extends to temporal dynamics analysis within individual speakers, Tier 3 advances to contextual inference and background understanding across multiple speakers, and Tier 4 culminates in comprehensive multi-speaker interaction understanding. Models can be assessed at each tier independently, enabling precise identification of strengths and limitations across the full spectrum of multi-speaker understanding tasks.

\paragraph{Tier 1: Single-Speaker Static Attribute Understanding.}
This tier focuses on identifying and characterizing individual speakers' static attributes. The primary objectives include speaker differentiation, demographic profiling (gender, age, accent), and paralinguistic analysis (voice quality, emotional tone). These capabilities establish the perceptual foundation necessary for higher-level reasoning tasks.

\paragraph{Tier 2: Single-Speaker Dynamic Attribute Understanding.}
Building upon static attribute recognition, this tier addresses temporal dynamics within individual speakers. Key capabilities include tracking emotional evolution, detecting voice quality changes, and identifying opinion shifts throughout conversations. Unlike Tier 1, this level requires models to reason about causality and context--understanding not just \emph{what} changes occur, but \emph{why} they occur. This tier evaluates models' ability to capture speaker-internal dynamics and infer cultural identity through language patterns and expression preferences.

\paragraph{Tier 3: Multi-Speaker Background Understanding.}
This tier focuses on contextual inference beyond immediate interaction dynamics. Tasks include inferring conversational venues, predicting dialogue outcomes, and determining speaker roles and social relationships. Models must analyze contextual cues, topic patterns, and language styles to understand the broader situational context. This tier evaluates the ability to reason about environmental factors and social structures that influence multi-speaker conversations.

\paragraph{Tier 4: Multi-Speaker Interaction Understanding.}
The highest tier extends analysis to inter-speaker dynamics and conversational interactions. Tasks include multi-speaker transcription with accurate attribution, paralinguistic and social interaction analysis, and cross-speaker reasoning about emotional shifts and motivations. This tier requires models to simultaneously track multiple speakers while reasoning about their mutual influence, conversational control, and collaborative behaviors. Success at this level demonstrates a comprehensive understanding of dynamic multi-party conversational structures.

The four-tier architecture ensures systematic progression from perceptual to reasoning tasks, from static to dynamic understanding, and from single-speaker to multi-speaker analysis. This design principle allows for granular evaluation of model capabilities while maintaining clear relationships between different complexity levels. 

\subsection{QA Construction Pipeline}
\begin{figure}[!htb]
\centering
\includegraphics[width=\columnwidth]{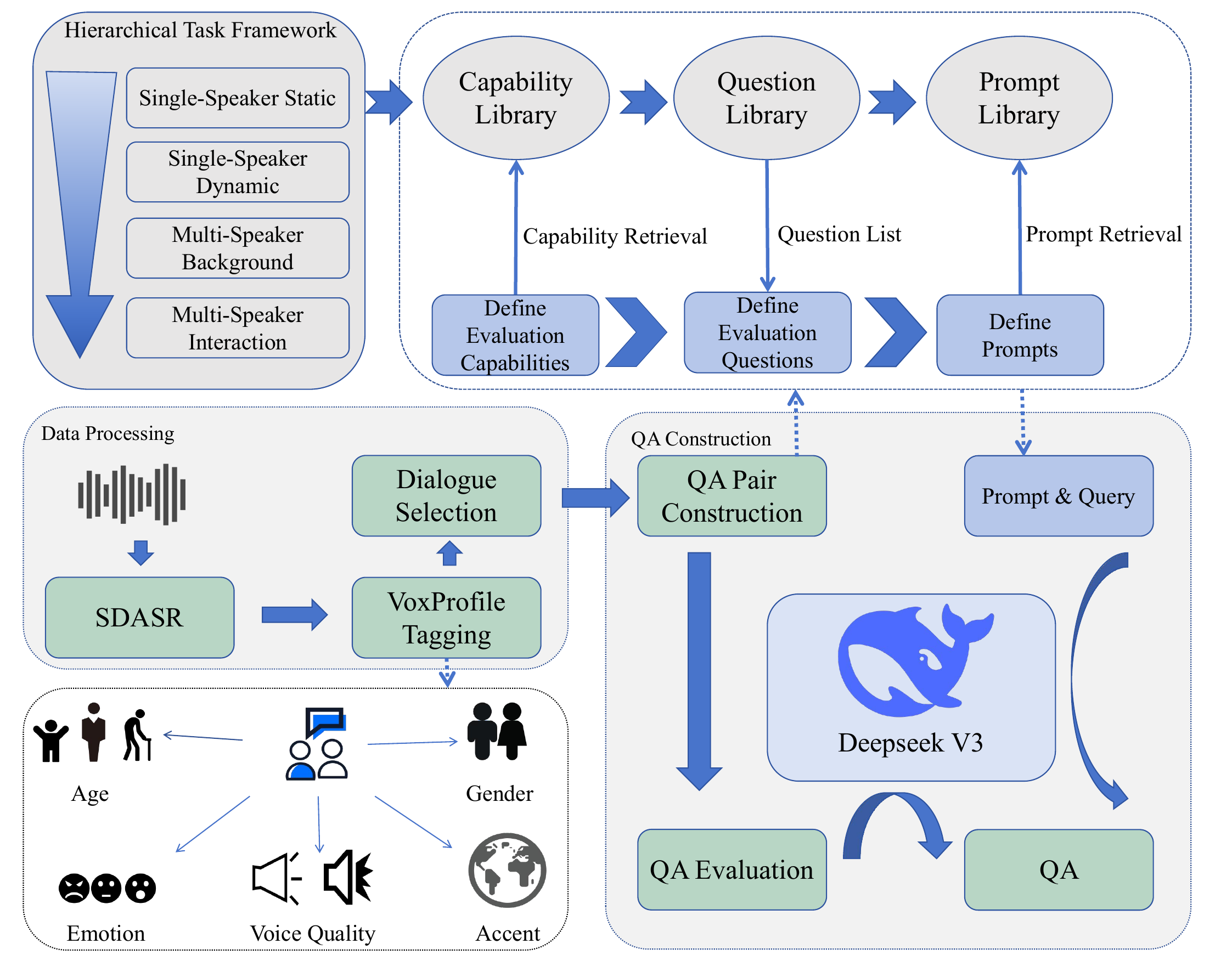}
\caption{The QA Construction Pipeline}
\label{fig:qapipe}
\end{figure}

To construct the four-tier benchmark with diverse speaker-centric audio-text tasks, we build a rigorous QA generation pipeline that automatically produces high-quality question–answer pairs from multi-speaker dialogues spanning various real-world scenarios and acoustic conditions. For each core ability, we design dedicated prompts to guide template construction and question formulation, ensuring that the resulting QA samples are tightly aligned with task-specific objectives. All selected audio segments span 60-120 seconds and include at least two speakers, thereby guaranteeing meaningful multi-speaker interaction and benchmark reliability. The overall QA pipeline(see Figure~\ref{fig:qapipe}) will be further detailed in the subsequent sections.

\paragraph{Data Selection and Preparation.}
To support the construction of multi-lingual and multi-scenario conversational datasets, we curate dialogue data from diverse real-world sources. For both near-field and far-field multi-speaker interactions, the full QA pipeline is applied to generate high-quality QA pairs. For film and television content, a denoising module was applied before subsequent processing. All QA data generated for downstream evaluation goes through an additional filtering stage, guided by large language model (LLM) assessments, to ensure its quality and relevance.

Specifically, the Chinese near-field data is sourced from \textit{MDT-AA007}, while the far-field data comes from \textit{alimeeting~\cite{Yu2022M2MeT,Yu2022Summary}}, a corpus of multi-speaker, meeting-style conversations recorded with distant microphones in acoustically complex environments.
For English, near-field data is collected from \textit{MDT-AD015}, which contains telephone-based conversational speech. The far-field data is drawn from \textit{CHiME6~\cite{watanabe2020chime}}, comprising real home-based multi-party conversations captured with distant microphones, featuring challenges such as background noise and overlapping speech.

\paragraph{Pipeline Design.}
Considering the complexity and heterogeneity of the data, the following sequential modules\footnote{The QA construction pipeline will be released alongside the benchmark} are employed, as demonstrated in Figure~\ref{fig:qapipe}:

\begin{enumerate}
    \item \textbf{Speaker Attributed Transcription:} Multi-speaker dialogues are first transcribed using an SDASR (Speaker-Diarized ASR) system, which aligns speech content with speaker identities over time.
    \item \textbf{Speaker Attribute Tagging:} The transcriptions are then processed by the \emph{VoxProfile} module, which annotates each speaker with metadata such as gender, role (e.g., host or guest), and secondary language usage.
    \item \textbf{Dialogue Segment Selection:} Utilizing Deepseek v3, dialogue segments are selected based on annotated speaker information and task-specific configuration, ensuring both relevance and diversity in the selected content.
    \item \textbf{QA Pair Construction:} QA pairs are generated from the selected segments using Deepseek v3, following predefined strategies tailored to the model's capabilities and the conversation context.
    \item \textbf{Automated QA Evaluation:} The resulting QA pairs undergo automated evaluation via Deepseek v3's built-in assessment mechanism, which measures QA quality against the original dialogue context and annotations.
\end{enumerate}

This structured and modular pipeline enables consistent, scalable, and high-quality QA data generation across complex, language-diverse, and acoustically challenging multi-speaker conversational scenarios.

\section{Experiments and Results}
\subsection{Dataset and Evaluation Protocols}

\subsubsection{Dataset}
MSU-Bench leverages six open-source datasets to cover diverse conversational scenarios~\footnote{Detailed statistics such as audio sources, duration distributions can be found in the appendix}: MDT-AA007 (Chinese telephone) and MDT-AD015 (English telephone) for near-field dialogue, AliMeeting~\cite{Yu2022M2MeT,Yu2022Summary} (Chinese meeting) and CHiME6~\cite{barker18_interspeech,watanabe2020chime} (English home dialogue) for far-field settings, and Chinese MovieClips and English MovieClips for challenging acoustic conditions (film audio post-processed to remove background music). For each task, sessions are randomly sampled from all datasets to ensure linguistic and acoustic diversity. Ten question-answer pairs per session are manually verified and constructed using varied templates, balancing phrasing diversity with consistent reasoning requirements. This ensures that performance differences are mainly attributable to audio complexity rather than variations in question form.
We find that multiple-choice questions can unintentionally provide LALMs with additional cues, potentially inflating performance. Instead, we use open-ended questions with answer formats and constraints in the inference prompts.

\subsubsection{Evaluation Protocols}
The evaluation uses a dedicated scoring prompt to assess LALM outputs along $3$ dimensions: relevance, accuracy, and causal soundness. Relevance ensures that responses are tightly aligned with questions, filtering out hallucinated or off-topic content. Accuracy measures the factual correctness of the response against the ground truth. Causal soundness evaluates the logical consistency of cause-effect reasoning in inference tasks; for non-causal questions, this component receives full marks by default.

\subsubsection{Speech Understanding Models}

This work evaluates five representative models\footnote{Due to input length restrictions (e.g., a 30-second audio limit), certain models could not be evaluated on multi-speaker scenarios. We also noticed the latest Step-Audio2~\cite{wu2025step} and Audio Flamingo3~\cite{goel2025audio}; however, as of submission, their model weights and inference code were not publicly available.} on MSU-Bench to assess multi-speaker understanding. Specifically, Gemini-2.5-Pro~\cite{team2023gemini}, Gemini-2.5-Flash~\cite{team2023gemini}, and GPT-4o-Audio are closed-source commercial systems, while Kimi-Audio~\cite{ding2025kimi} and Qwen2.5-Omni~\cite{xu2025qwen2} are open-source models. This diverse selection enables a systematic comparison between open-source and commercial solutions across all benchmark tiers and tasks.

\subsection{Evaluation Results and Analysis}
\begin{figure}[!htb]
\centering
\includegraphics[width=0.48\textwidth]{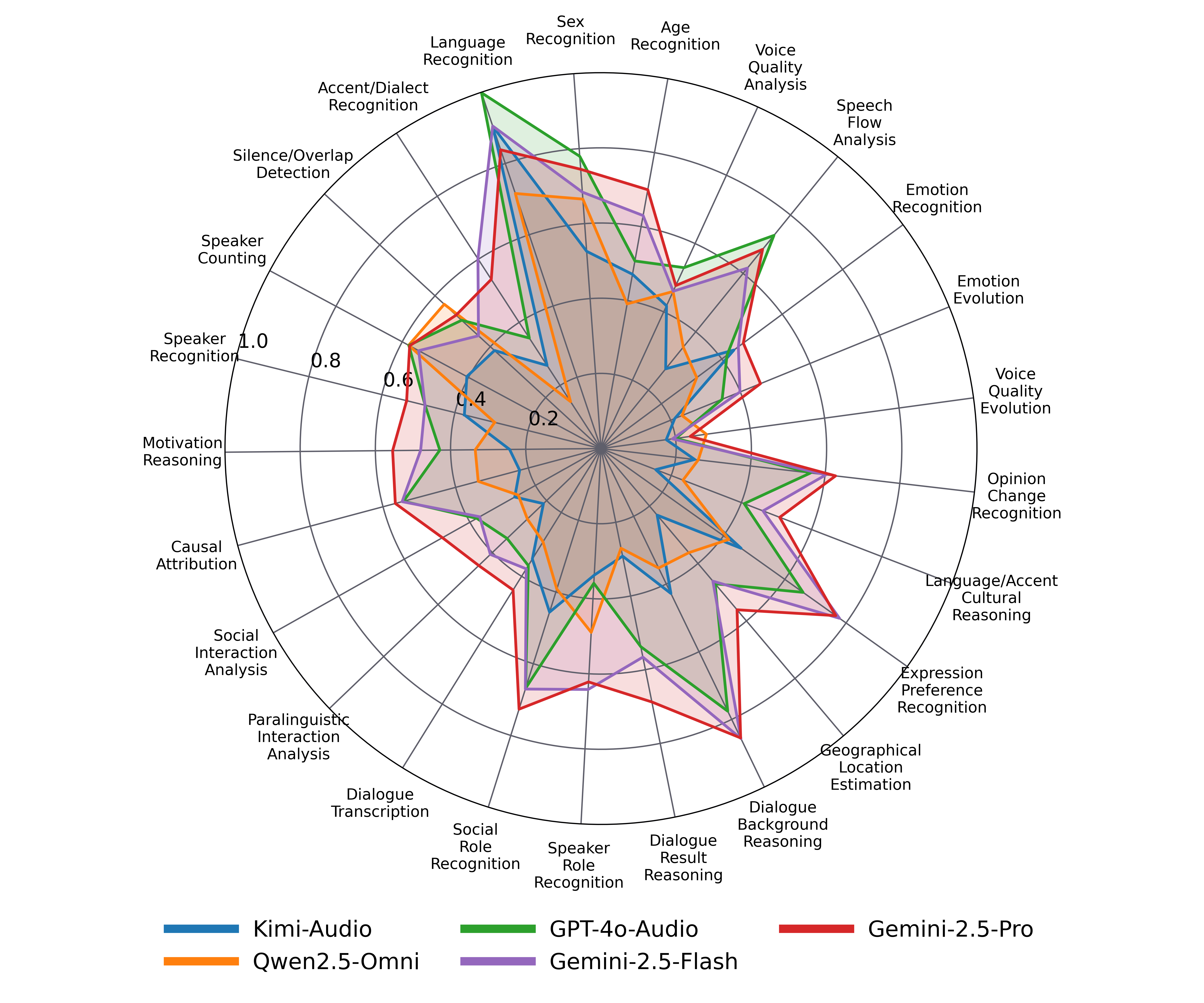}
\caption{Overall Performance Comparison}
\label{fig:pcom}
\vspace{-4mm}
\end{figure}

We present comprehensive evaluation results of both state-of-the-art open-source and commercial models on our benchmark (see Table~\ref{tab:overall_results}). The test set is carefully balanced, with samples drawn uniformly from six diverse data sources. For each benchmark tier and capability, we report the average results of all tasks related to that capability, with the task-capability mappings detailed in Table~\ref{tab:tiersep}. Moreover, the comparison of different systems across all 25 tasks is illustrated in Figure~\ref{fig:pcom}, providing a highly intuitive performance comparison across models.

\paragraph{Commercial vs. Open-source Performance Gap.} Our evaluation reveals a significant and consistent performance gap between commercial and open-source models. The Gemini series shows superior performance across all tiers, with Gemini-2.5-Pro achieving the best results in 8 out of 9 capabilities. GPT-4o-Audio shows strong performance in paralinguistic analysis (0.55 in SPA) but weaker performance in cultural identity integration (0.52 in SCII). Among open-source models, Kimi-Audio shows relatively strong performance in speaker recognition (0.39 in SR) but struggles significantly with multi-speaker interaction analysis (0.23 in MSIA), indicating limitations in handling complex multi-party dynamics. This suggests that the involved commercial models have significantly better capabilities in handling complex multi-speaker reasoning tasks.

\subsubsection{Static vs. Dynamic Attribute Understanding}
Our analysis reveals fundamental differences in how models handle static versus dynamic speaker attributes. Static attributes (e.g., age, gender) remain constant in dialogue, while dynamic ones (e.g., emotion, prosody) vary with context. To bridge the training–inference label gap in audio models, we provide explicit labels during inference. Evaluation shows large performance gaps across models: Gemini~\cite{team2023gemini} excels in age, gender, and accent recognition, while GPT-4o is accurate on gender but weaker on age and accent. Most models struggle with dynamic attributes like timbre and emotion, revealing limitations in current LALMs' paralinguistic understanding.

\subsubsection{Acoustic vs. Semantic Processing Patterns}
Our results reveal a clear preference for semantic over acoustic processing across all models. Tasks that primarily rely on semantic content (e.g., SAC, MSSI) achieve significantly higher performance than those requiring acoustic analysis (e.g., SPA, SDA). This pattern is consistent across all tiers and model families, suggesting a fundamental limitation in current LALM architectures.
This pattern suggests that current LALMs can not perfectly leverage fine-grained acoustic features, even when these features are explicitly relevant to the task. This limitation has significant implications for real-world applications where acoustic cues are crucial for understanding speaker intent and emotional state.

\subsubsection{Cross-Lingual Performance Analysis}
\begin{figure}[!htb]
\centering
\includegraphics[width=0.47\textwidth]{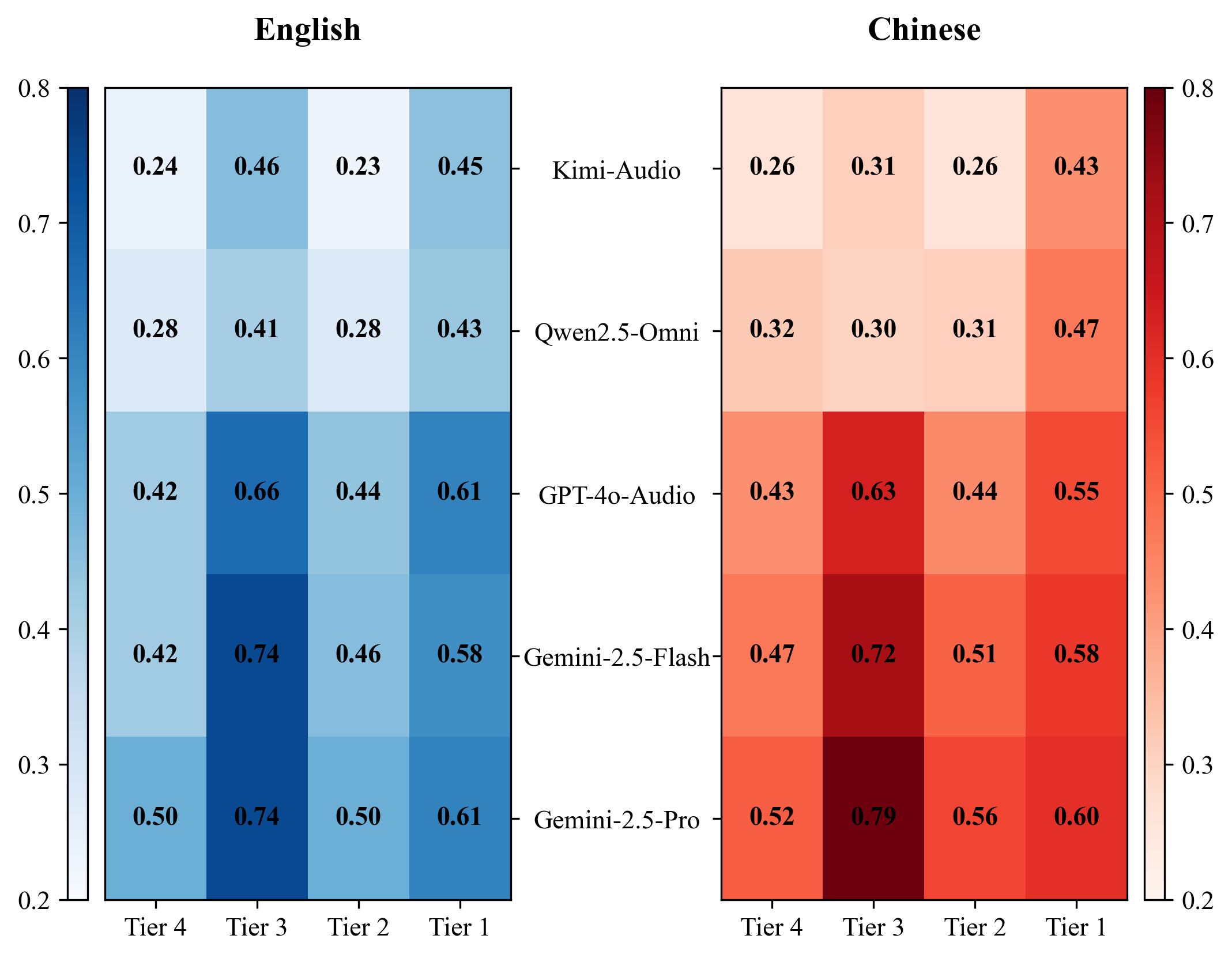}
\caption{Performance comparison of different models on tasks of different languages}
\label{fig:language}
\vspace{-4mm}
\end{figure}

MSU-Bench enables natural cross-lingual evaluation, revealing interesting patterns in model performance across English and Chinese. We present the average results of different tier tasks for each model through a heatmap visualization, as shown in Figure~\ref{fig:language}. The results demonstrate that different models exhibit largely consistent trends across both English and Chinese languages, indicating that these models do not exhibit over-optimization for any specific language. This also demonstrates that the difficulty levels of our selected English and Chinese data sources are comparable.

\subsubsection{Paralinguistic context v.s. Semantic context}

Our analysis of Tier 3 (Multi-Speaker Background Understanding) versus Tier 4 (Multi-Speaker Interaction Understanding) reveals important distinctions in how models process different types of contextual information. Tier 3 tasks focus on background understanding and rely heavily on semantic content to infer dialogue scenes and speaker relationships. Notably, even with speaker attribution errors, models with transcription capability can still identify the dialogue setting and social relations based on topical cues. In contrast, Tier 4 tasks require deeper analysis of emotional and semantic exchanges, serving as a more sophisticated analysis of speaker dynamics. When emotional shifts are explicitly marked in QA settings, models leveraging transcription and semantic reasoning can infer causes accurately. However, when inferring emotions solely from dialogue segments, models often accumulate errors due to the difficulty of tracking emotional dynamics across multiple speakers.

\subsection{Error Analysis}

For a systematic comparison of error causes, we select Gemini-2.5-Pro and Kimi-Audio as two representatives of commercial and open-source models, aggregating results across all 25 sub-tasks and following MMSU's taxonomy to classify errors into several key categories. We randomly sampled 200 QA pairs from error cases across different tiers for both Gemini-2.5-Pro and Kimi-Audio and analyzed the error causes, as summarized in Table~\ref{tab:error_distribution}. 

\begin{table}[ht]
\centering
\adjustbox{width=\columnwidth,center}
{%
\renewcommand{\arraystretch}{1.2}
\setlength{\tabcolsep}{8pt}
\footnotesize
\rowcolors{2}{gray!8}{white}
\begin{tabular}{lcc}
\toprule
\textbf{Error Type} & \textbf{Gemini-2.5-Pro (\%)} & \textbf{Kimi-Audio (\%)} \\
\midrule
Rejection of Answer      & -- & 0.50 \\
Answer Extraction Errors & 4.04 & 48.24 \\
Perceptual Errors        & 56.06 & 37.68 \\
Reasoning Errors         & 37.37 & 13.57 \\
\rowcolor{gray!15}
Lack of Knowledge        & 2.53 & -- \\
\bottomrule
\end{tabular}
}
\caption{Error distribution analysis}
\label{tab:error_distribution}
\end{table}
\vspace{-2mm}

Our evaluation reveals that Gemini-2.5-Pro provides comprehensive responses with strong instruction-following, while Kimi-Audio frequently delivers partial answers. Both models exhibit significant perception errors, consistent with MMSU findings. Additional analysis of Tier 4 error distribution shows perception errors dominate at 68.09\%, highlighting the necessity of strong multi-speaker perception for effective interaction comprehension\footnote{Tier-wise error distribution can be found in the appendix}.

\section{Conclusion}
We present MSU-Bench, a comprehensive four-tier benchmark for multi-speaker conversational understanding that systematically evaluates speaker-centric capabilities from basic perception to complex reasoning. Our hierarchical framework covers single-speaker static/dynamic attribute understanding and multi-speaker background/interaction understanding, with all tasks grounded in authentic conversational contexts. Through extensive evaluation of state-of-the-art models, we demonstrate significant performance degradation as task complexity increases across tiers, revealing persistent gaps between open-source and commercial solutions, particularly in multi-speaker interaction reasoning. Our analysis reveals critical limitations in current LALMs' ability to handle fine-grained acoustic cues and complex multi-speaker dynamics, highlighting the need for more efforts towards achieving conversational understanding. MSU-Bench provides a standardized evaluation platform to facilitate future research in multi-speaker speech understanding and guide the development of more robust conversational AI systems.

\noindent\textbf{Limitations:} Despite efforts to diversify scenarios and models, access limitations may constrain comprehensiveness. We hope this dataset helps identify audio-language model performance on speaker-centric tasks.

\bibliography{aaai2026}
\clearpage
\begin{center}
    {\Huge \bfseries Appendix}
\end{center}
\vspace{1em}

\section{Data Construction}
This section details the datasets and procedures used to construct our multi-speaker audio understanding benchmark.

\subsection{Data Sources}
We selected six publicly available datasets, three in English and three in Chinese, to cover a wide range of conversational scenarios, acoustic conditions, and speaker behaviors.

\subsubsection{English Datasets}
\begin{itemize}
    \item \textbf{MDT-AD015} is an open-source Magic Data corpus of dual-speaker, close-talk English telephone conversations. The speech is highly conversational, featuring frequent interruptions, backchannels, and emotional expressions. The original 8kHz audio, which contains channel loss and background noise, was upsampled to 16kHz. We used the entire 5-hour open-source portion for QA generation, selecting only the first audio channel.

    \item \textbf{CHiME-6} contains distant-microphone, multi-speaker conversations from naturalistic home environments. The dataset captures spontaneous dinner-time interactions. From the 32-hour training set, we randomly sampled 6 hours of audio (from 3 sessions) for QA generation. The audio is provided at a 16kHz sampling rate, and we used the first channel for our experiments.

    \item \textbf{en-film} is a collection of English movie and television dialogues. This dataset features acoustically complex scenes with two or more speakers in both formal and informal settings, and includes channel distortion, ambient noise, and background music. After an initial transcription by an ASR model, we manually filtered the data to obtain approximately 41 hours of dialogue segments suitable for audio reasoning tasks, which were then used for QA generation.
\end{itemize}

\subsubsection{Chinese Datasets}
\begin{itemize}
    \item \textbf{MDT-AA007} is an open-source Magic Data corpus of dual-speaker Mandarin Chinese telephone conversations. It features natural, close-talk conversational speech with typical channel degradation and slight background noise. The public release includes 15 sessions, totaling 5.2 hours, all of which were used in our work.

    \item \textbf{AliMeeting} is a far-field, multi-speaker Mandarin meeting dataset recorded in formal settings. It is characterized by complex acoustic conditions, including speech overlap, interruptions, noise, and room reverberation. We randomly selected 8 sessions (4 hours) for QA generation. The audio is sampled at 16kHz, and we utilized the first channel.

    \item \textbf{cn-Film} comprises Mandarin Chinese movie and television dialogues from diverse and rich acoustic environments that include noise, reverberation, sound events, and background audio. The data exhibits frequent conversational phenomena such as interruptions and overlapping speech, making it highly suitable for generating challenging test cases.
\end{itemize}

\subsection{Two-stage QA Construction}

 Our benchmark employs a two-stage construction approach: first, we develop an automated pipeline to generate large-scale candidate QA pairs from six diverse datasets; second, we implement rigorous filtering and selection procedures to finalize the benchmark. 
 
\paragraph{General Design Philosophy and Automated Pipeline}
\begin{itemize}

\item \textbf{Speaker-Centric Question Design.} Our benchmark distinguishes itself from existing evaluations through its focus on open-domain question answering for dialogue understanding. We adopt a speaker-centric approach by formulating diverse question templates that comprehensively assess model capabilities across multiple dimensions. To ensure evaluation precision, we provide contextual dialogue information that enables models to identify relevant content accurately, thereby guaranteeing that each question yields a single, unambiguous correct answer. This design principle eliminates potential ambiguities inherent in broad questions such as "What is the relationship between the speakers?", which could generate multiple valid interpretations depending on whether pairwise or collective relationships are being assessed.

\item \textbf{Automated QA Generation Pipeline.} Our QA generation pipeline employs capability-specific prompts that guide the creation of targeted question-answer pairs. For each task within a given capability, we establish explicit construction methodologies and template requirements. We utilize placeholder tokens (e.g., \texttt{<spkid>}) that enable large language models to autonomously identify appropriate dialogue segments and construct contextually relevant QAs, leveraging their inherent reasoning capabilities. This methodology enhances question diversity while maintaining quality control by preventing the generation of semantically invalid or contextually inappropriate questions.

\item \textbf{Audio Sample Constraints.} The pipeline architecture supports scalable generation of task-specific QA pairs that maintain both accuracy and alignment with evaluation objectives. We constrain audio samples to durations of 60-120 seconds and enforce a minimum threshold of two speakers per sample to ensure sufficient conversational complexity for meaningful evaluation.

\end{itemize}

\paragraph{Filtering and Selection Process}
\begin{itemize}
\item \textbf{Benchmark Curation and Sampling.} Our automated pipeline generates a large candidate pool of QA pairs by drawing from six datasets that span diverse linguistic and acoustic environments. To curate the final benchmark from this pool, we sample representative QA examples for each evaluation task, ensuring broad coverage while maintaining efficiency. Every selected QA pair undergoes rigorous manual verification to guarantee its quality before inclusion in the benchmark.

\item \textbf{Template Consistency Control.} Given our multi-template design approach, individual tasks may exhibit varying question formulations during the sampling process. We maintain strict quality control by ensuring that all question templates within a single task assess identical underlying capabilities and preserve consistent difficulty levels. This design ensures that model performance variations reflect the inherent complexity of input conversations rather than artifacts of question phrasing or template selection.

\item \textbf{Question Format Selection.} During QA construction, we identified that multiple-choice formats could inadvertently provide large audio-language models with supplementary conversational cues, potentially introducing evaluation bias that obscures genuine dialogue understanding capabilities. To mitigate this concern, we exclusively employ open-ended QA templates while embedding format specifications and scope constraints within inference prompts (\textbf{All prompts can be found at the end of this appendix.})
, thereby directing models to respond according to precise task requirements without external contextual influences.
\item \textbf{Quality Assurance}
We implement a multi-stage quality control process in which large language models perform initial filtering to eliminate substandard samples, establishing a foundation for benchmark quality. For reasoning-intensive questions, which are susceptible to annotation errors or involve complex inference processes, we conduct a comprehensive manual review and correction to ensure evaluation accuracy and maintain benchmark integrity.
\end{itemize}

\begin{figure}[!htb]
\centering
\includegraphics[width=0.47\textwidth]{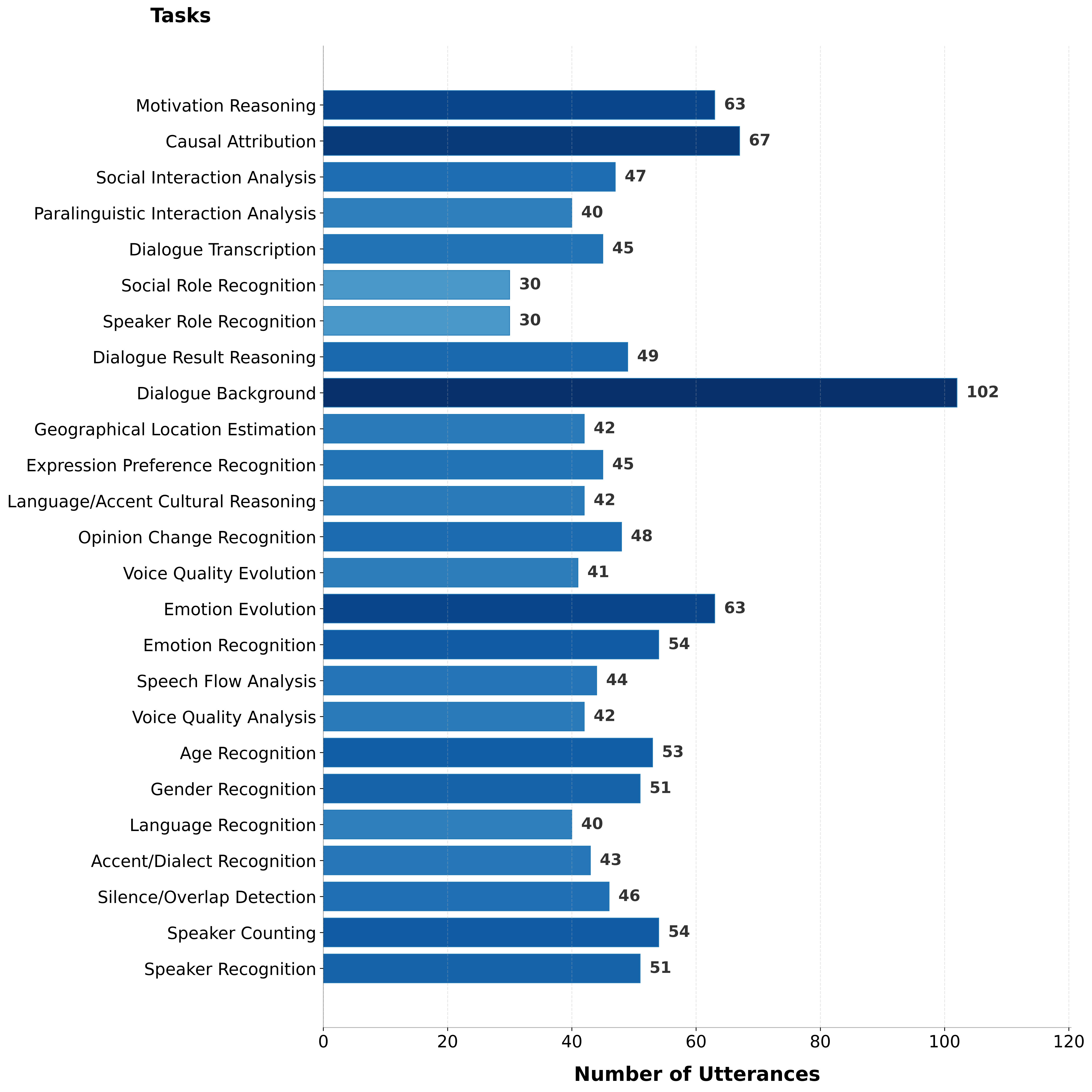}
\caption{Number of QA pairs for different tasks in MSU-Bench}
\label{fig:count}
\end{figure}

After the filtering and selection process, our final benchmark comprises 25 tasks totaling 1232 questions. The detailed distribution across different capabilities and tasks is illustrated in Figure~\ref{fig:count}.

\subsection{Evaluation Pipeline}
\paragraph{Infer Protocol}
To ensure consistent response formatting and avoid potential mismatches between training and inference labels, we design dedicated inference prompts for LALM-based reasoning. The inference prompt examples are presented in Figure~\ref{fig:inference_frame}, along with the task description, input/output specifications, subtask inference requirements, and other constraints. Detailed specifications for subtask inference are provided in Figures~\ref{fig:inference_speaker_transcription}--\ref{fig:inference_multi_speaker_interaction_understanding}. These prompts explicitly specify the expected response format and key focus points according to each task and question template. For example, multi-speaker transcription tasks require responses in the structured format \texttt{\{spk\_1:$\cdots$ ;spk\_2: $\cdots$ ;\}}, while age or emotion classification tasks provide predefined candidate labels for appropriate model selection.
\paragraph{Evaluation Protocol}
When applying an LLM to score LALM responses, we construct separate scoring prompts tailored to each task. The evaluation criteria include response relevance (including detection of hallucinations and irrelevant multi-hop reasoning), answer correctness, and logical consistency (particularly examining the accuracy of inferred causal relationships in reasoning tasks). An example of the evaluation prompt is presented in Figure~\ref{fig:evaluation_frame}, accompanied by a detailed explanation of the scoring procedure. The model first categorizes the questions into parallel and progressive types. It then performs a three-dimensional evaluation based on the decomposed questions and their specific scoring criteria. Finally, an overall score is produced. The detailed requirements of sub-task evaluation are provided in Figures~\ref{fig:evaluation_speaker_counting}--\ref{fig:evaluation_dialogue_outcome_reasoning}. This multidimensional assessment framework ensures a comprehensive review of model capabilities across diverse dialogue.

\section{Task Examples}

While Table 2 in the main text presents the 10 capabilities and their corresponding 25 tasks, we provide concrete examples for each task in the following Table~\ref{tab:task_examples} to facilitate readers' understanding of the specific content and requirements of individual evaluation tasks.

\begin{table*}[!ht]
\centering
\begin{tabular}{@{}p{0.15\textwidth}p{0.15\textwidth}p{0.25\textwidth}p{0.36\textwidth}@{}}
\toprule
\textbf{Tier} & \textbf{Ability} & \textbf{Task} & \textbf{Sample QA} \\
\midrule

\multirow{10}{0.15\textwidth}{\textbf{Tier 1:} Single-Speaker Static Attribute Understanding} & 
\multirow{3}{0.15\textwidth}{Speaker Recognition} & 
Speaker Recognition & 
\textbf{Q:} What are all utterances made by the second speaker in the audio? \textbf{A:} "I know. But I still have to survive..." \\
\cmidrule{3-4}
& & Speaker Counting & \textbf{Q:} How many different speakers are in this recording? \textbf{A:} Two. \\
\cmidrule{3-4}
& & Silence/Overlap Detection & \textbf{Q:} Is Speaker\_1 interrupted after saying "That's why I came to find you..."? \textbf{A:} Yes, Speaker\_2 interrupted... \\

\cmidrule{2-4}
& \multirow{4}{0.15\textwidth}{Speaker Attribute Comprehension} & 
Accent/Dialect Recognition & 
\textbf{Q:} What accent does Speaker\_1 have when saying "Uncle Ding..."? \textbf{A:} Speaker\_1 has a Beijing accent. \\
\cmidrule{3-4}
& & Language Recognition & \textbf{Q:} What language(s) are used in the audio? \textbf{A:} Chinese. \\
\cmidrule{3-4}
& & Gender Recognition & \textbf{Q:} Is Speaker\_1 male or female? \textbf{A:} Male. \\
\cmidrule{3-4}
& & Age Recognition & \textbf{Q:} What is Speaker\_1's age group? \textbf{A:} Young adult. \\

\cmidrule{2-4}
& \multirow{3}{0.15\textwidth}{Speaker Paralinguistic Analysis} & 
Voice Quality Analysis & 
\textbf{Q:} How does the volume of Speaker\_1's voice sound when saying "Uncle Ding..."? \textbf{A:} Authoritative. \\
\cmidrule{3-4}
& & Speech Flow Analysis & \textbf{Q:} Did Speaker\_2 have any pauses while expressing "I know..."? \textbf{A:} He paused while saying "I know". \\
\cmidrule{3-4}
& & Emotion Recognition & \textbf{Q:} What is Speaker\_2's emotion when saying "I know..."? \textbf{A:} Sad. \\

\midrule
\multirow{6}{0.15\textwidth}{\textbf{Tier 2:} Single-Speaker Dynamic Attribute Understanding} & 
\multirow{3}{0.15\textwidth}{Speaker Dynamic Analysis} & 
Emotion Evolution & 
\textbf{Q:} How does Speaker\_2's emotion evolve during the dialogue? \textbf{A:} Speaker\_2 sounds sad when saying "I know..." then happy when saying "Ah, sure..." \\
\cmidrule{3-4}
& & Voice Quality Evolution & \textbf{Q:} Does Speaker\_2's volume change from nasal to shrill? \textbf{A:} Speaker\_2's pitch is nasal when saying "Ah, sure..." and shrill when saying "Sweep the floor?" \\
\cmidrule{3-4}
& & Opinion Change Recognition & \textbf{Q:} Does Speaker\_1's main concern change? \textbf{A:} Speaker\_1 goes from warning to requesting help to issuing threats. \\

\cmidrule{2-4}
& \multirow{3}{0.15\textwidth}{Speaker Cultural Identity Integration} & 
Language/Accent Cultural Reasoning & 
\textbf{Q:} What is Speaker\_3's accent and how does it relate to their viewpoint? \textbf{A:} Speaker\_3 has a Taiwanese accent, mentioning Twitter access issues in mainland China... \\
\cmidrule{3-4}
& & Expression Preference Recognition & \textbf{Q:} What age group is Speaker\_3 most likely in? \textbf{A:} Young adult, based on internet and social media topics. \\
\cmidrule{3-4}
& & Geographical Location Estimation & \textbf{Q:} Who is likely to be from Taiwan? \textbf{A:} Speaker\_3, based on Taiwanese accent and regional expressions. \\
\midrule

\end{tabular}
\end{table*}
\clearpage
\begin{table*}[!ht]
\centering
\begin{tabular}{@{}p{0.15\textwidth}p{0.15\textwidth}p{0.25\textwidth}p{0.36\textwidth}@{}}

\midrule
\multirow{4}{0.15\textwidth}{\textbf{Tier 3:} Multi-Speaker Background Understanding} & 
\multirow{2}{0.15\textwidth}{Multi-Speaker Scene Inference} & 
Dialogue Background & 
\textbf{Q:} Is this conversation formal or casual? \textbf{A:} Semi-formal daily life setting with serious topics but colloquial expressions. \\
\cmidrule{3-4}
& & Dialogue Result Reasoning & \textbf{Q:} Does this conversation lead to consensus or disagreement? \textbf{A:} Consensus through Speaker\_1's dominant position and implied threats. \\

\cmidrule{2-4}
& \multirow{2}{0.15\textwidth}{Multi-Speaker Relationship Inference} & 
Speaker Role Recognition & 
\textbf{Q:} Does Speaker\_1 play a dominant role? \textbf{A:} Yes, Speaker\_1 dominates while Speaker\_2 remains passive. \\
\cmidrule{3-4}
& & Social Role Recognition & \textbf{Q:} What is the relationship between speakers? \textbf{A:} Superior-subordinate based on authoritative tone and command patterns. \\

\midrule
\multirow{5}{0.15\textwidth}{\textbf{Tier 4:} Multi-Speaker Interaction Understanding} & 
Multi-Speaker Transcription & 
Dialogue Transcription & 
\textbf{Q:} What does each speaker say? \textbf{A:} spk\_1: Uncle Ding, you can't contact your sister... spk\_2: I know. But I still have to survive... \\

\cmidrule{2-4}
& \multirow{2}{0.15\textwidth}{Multi-Speaker Interaction Analysis} & 
Paralinguistic Interaction Analysis & 
\textbf{Q:} What is Speaker\_2's emotion when saying "Sweep the floor?" \textbf{A:} Surprise, which made Speaker\_1 dissatisfied and angry. \\
\cmidrule{3-4}
& & Social Interaction Analysis & \textbf{Q:} What did Speaker\_2 say to express agreement? \textbf{A:} "Ah, sure, sure. My English is pretty good..." \\

\cmidrule{2-4}
& \multirow{2}{0.15\textwidth}{Multi-Speaker Contextual Reasoning} & 
Causal Attribution & 
\textbf{Q:} What are Speaker\_2's emotions and are they related to Speaker\_1's words? \textbf{A:} Sadness then happiness, related to Speaker\_1's suggestion. \\
\cmidrule{3-4}
& & Motivation Reasoning & \textbf{Q:} What did Speaker\_2 do when hearing about the favor? \textbf{A:} Interrupted immediately, expressing eagerness to leave the hotel. \\

\bottomrule
\end{tabular}
\caption{Task examples across different tiers and abilities in MSU-Bench}
\label{tab:task_examples}
\end{table*}

\clearpage

\section{Additional Results and Analysis}

\subsection{Tier-wise Error Analysis}

Table 4 in the main text delineates the error types and is exemplified with corresponding samples per error type from the responses of Gemini 2.5 Pro and Kimi-Audio.
To further investigate the causes of errors at each tier, Table~\ref{tab:tier_error} presents the distribution of error types across tiers.

\begin{table}[!htb]
\centering
\renewcommand{\arraystretch}{1.35}
\small
\begin{tabular}{@{}>{\raggedright}p{0.05\textwidth}>{\raggedright}p{0.19\textwidth}>{\centering}p{0.08\textwidth}>{\centering\arraybackslash}p{0.08\textwidth}@{}}
\toprule
\multirow{2}{*}{\textbf{Tier}} & \multirow{2}{*}{\textbf{Error Type}} & \multicolumn{2}{c}{\textbf{Error Distribution (\%)}} \\ 
\cmidrule(lr){3-4}
& & \textbf{Gemini-2.5-Pro} & \textbf{Kimi-Audio} \\ 
\midrule
\textbf{Tier 1} & Rejection of Answer & -- & -- \\
& Answer Extraction Errors & 10.77 & 22.58 \\
& Perceptual Errors & 81.54 & 77.42 \\
& Reasoning Errors & 7.69 & -- \\
& Lack of Knowledge & -- & -- \\ \midrule
\addlinespace[0.2cm] 

\textbf{Tier 2} & Rejection of Answer & -- & -- \\
& Answer Extraction Errors & 1.67 & 68.97 \\
& Perceptual Errors & 43.33 & 22.41 \\
& Reasoning Errors & 46.67 & 8.62 \\
& Lack of Knowledge & 8.33 & -- \\\midrule
\addlinespace[0.2cm]

\textbf{Tier 3} & Rejection of Answer & -- & 3.23 \\
& Answer Extraction Errors & -- & 67.74 \\
& Perceptual Errors & -- & -- \\
& Reasoning Errors & 100 & 29.03 \\
& Lack of Knowledge & -- & -- \\ \midrule
\addlinespace[0.2cm]

\textbf{Tier 4} & Rejection of Answer & -- & -- \\
& Answer Extraction Errors & -- & 43.75 \\
& Perceptual Errors & 68.09 & 29.17 \\
& Reasoning Errors & 31.91 & 27.08 \\
& Lack of Knowledge & -- & -- \\
\bottomrule
\end{tabular}
\caption{Error distribution at different tiers}
\label{tab:tier_error}
\end{table}

Tier 1 focuses on recognizing and understanding the static attributes of speakers, with Perceptual Errors being its predominant error type. Notably, Tier 1 also contains Reasoning Errors. These arise when Tier 1 tasks require reasoning based on the integration of multi-dimensional speaker attributes (e.g., the silence/overlap detection task necessitates combining speaker turns and temporal information). Though the model correctly understands the audio content, it fails to adequately integrate relevant information, resulting in Reasoning Errors.
Furthermore, our analysis of Gemini 2.5 Pro reveals that Tier 3 Reasoning Errors constitute 100\%. This predominance is primarily due to the nature of Tier 3, which emphasizes multi-speaker contextual understanding. These tasks require models to integrate semantic information from multiple speakers and perform in-depth reasoning to infer conversational elements such as scenarios, identities, and relationships. Consequently, reasoning errors emerge as the primary error source under this tier, particularly for advanced audio understanding models that already exhibit strong semantic perception capabilities.

Following the error type definitions established in the MMSU paper, we categorize and analyze model errors accordingly in the main text. To facilitate reader comprehension, Table~\ref{tab:error_definition} provides detailed explanations of each error type along with illustrative examples from our evaluation

\subsection{Model Performance Analysis}
\begin{figure}[htb]
\centering
\begin{tabular}{cc}
\includegraphics[width=0.45\columnwidth]{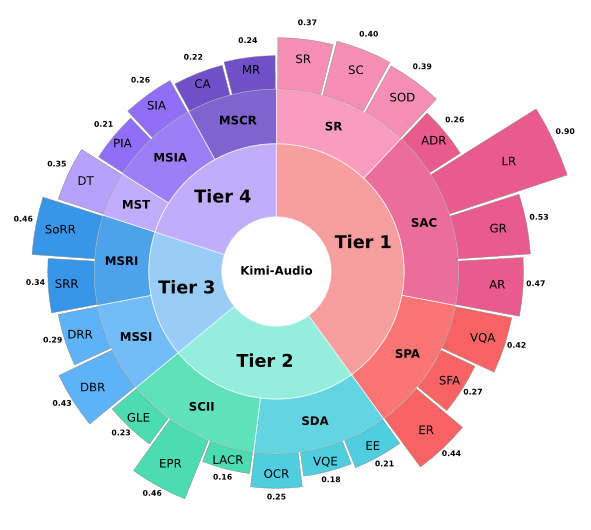} &
\includegraphics[width=0.45\columnwidth]{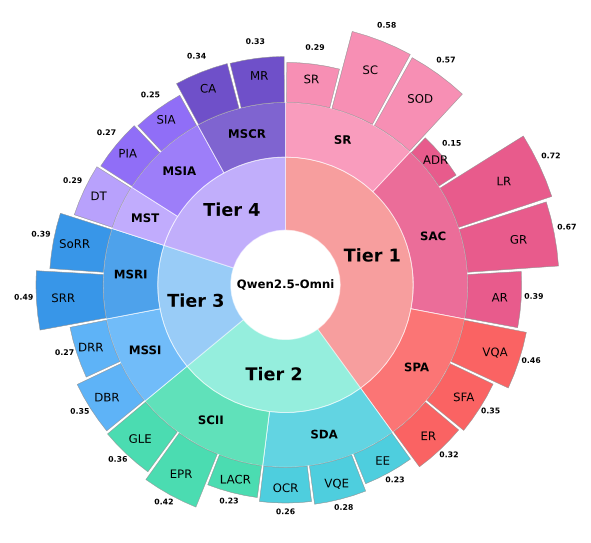} \\
\includegraphics[width=0.45\columnwidth]{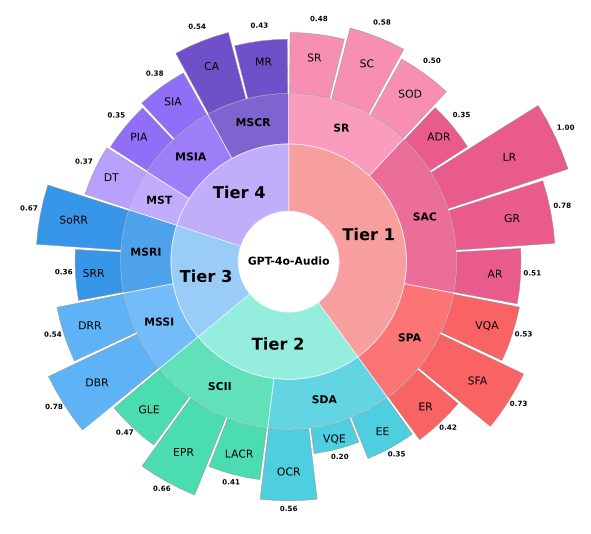} &
\includegraphics[width=0.45\columnwidth]{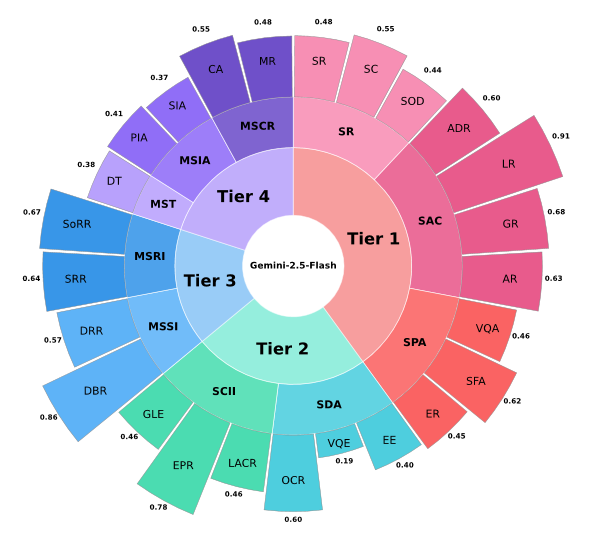} \\
\multicolumn{2}{c}{\includegraphics[width=0.45\columnwidth]{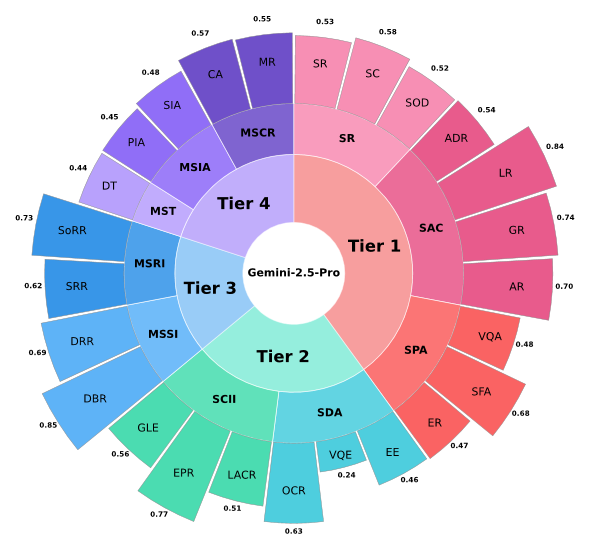}}
\end{tabular}
\caption{Task-wise performance breakdown for each evaluated model. Each subplot displays the accuracy scores across all 25 tasks for: (top row) Kimi Audio and Qwen2.5-Omni, (middle row) GPT-4o-Audio and Gemini-2.5-Flash, (bottom) Gemini-2.5-Pro.}
\label{fig:ind_model}
\end{figure}

\begin{figure*}[!htb]
\includegraphics[width=1.0\textwidth]{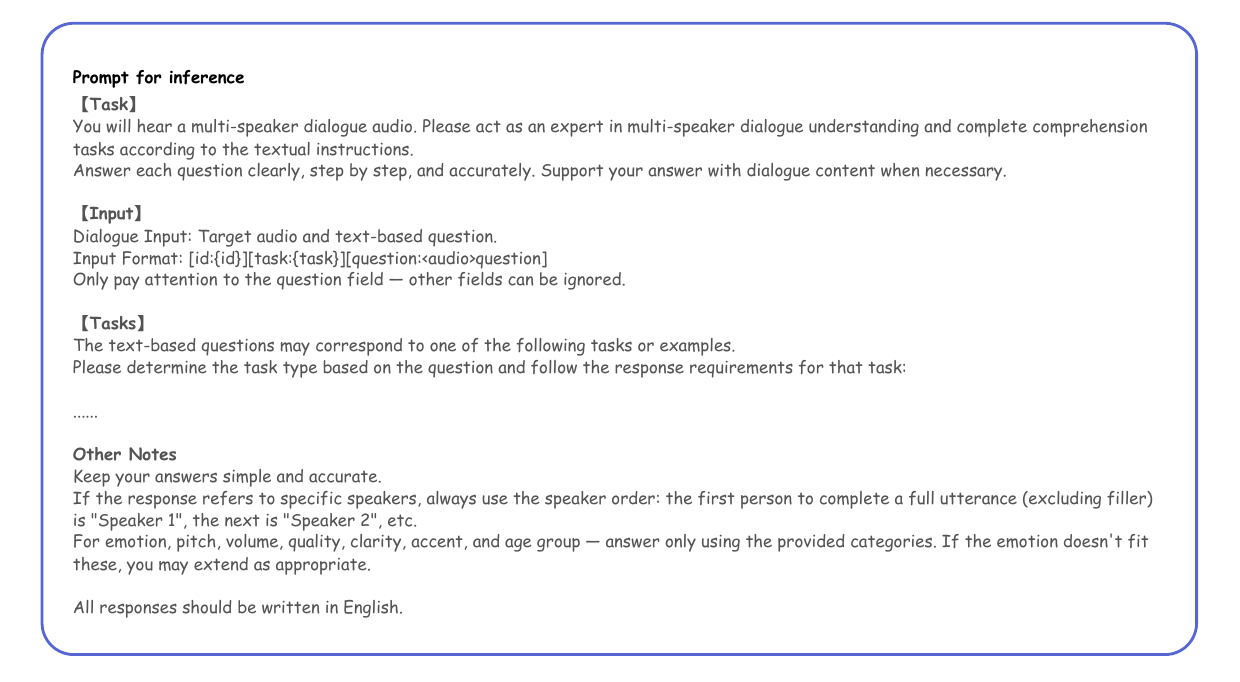}
\caption{LALM Inference Prompt}
\label{fig:inference_frame}
\end{figure*}

\begin{figure*}[!htb]
\includegraphics[width=1.0\textwidth]{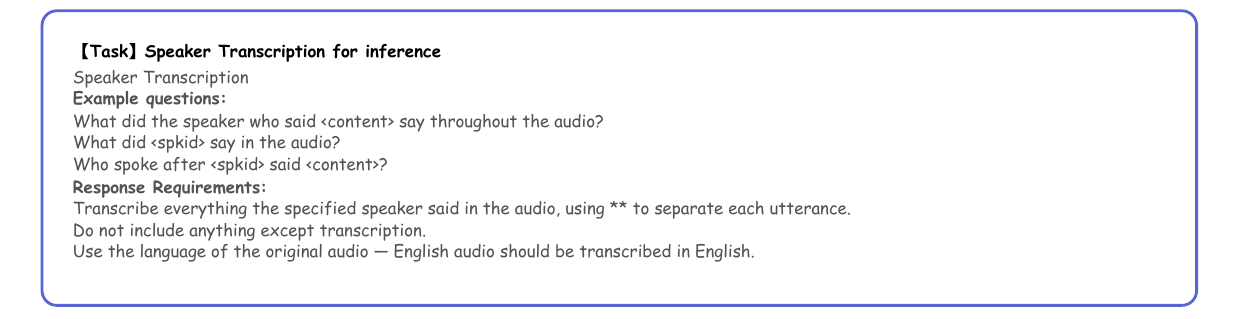}
\caption{Example Questions and Requirements for Speaker Tranription Task}
\label{fig:inference_speaker_transcription}
\end{figure*}

\begin{figure*}[!htb]
\includegraphics[width=1.0\textwidth]{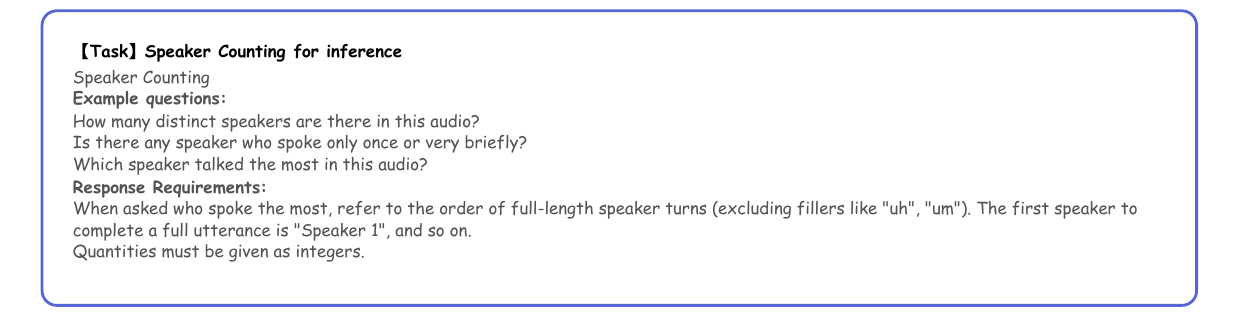}
\caption{Example Question and Requirements for Speaker Counting}
\label{fig:inference_speaker_counting}
\end{figure*}

\begin{figure*}[!htb]
\includegraphics[width=1.0\textwidth]{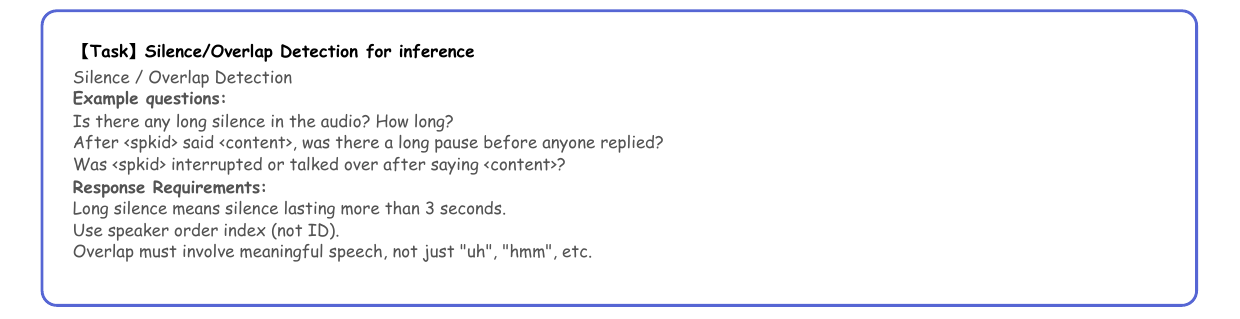}
\caption{Example Question and Requirements for Silence/Overlap Detection}
\label{fig:inference_silence_overlap_detection}
\end{figure*}

\begin{figure*}[!htb]
\includegraphics[width=1.0\textwidth]{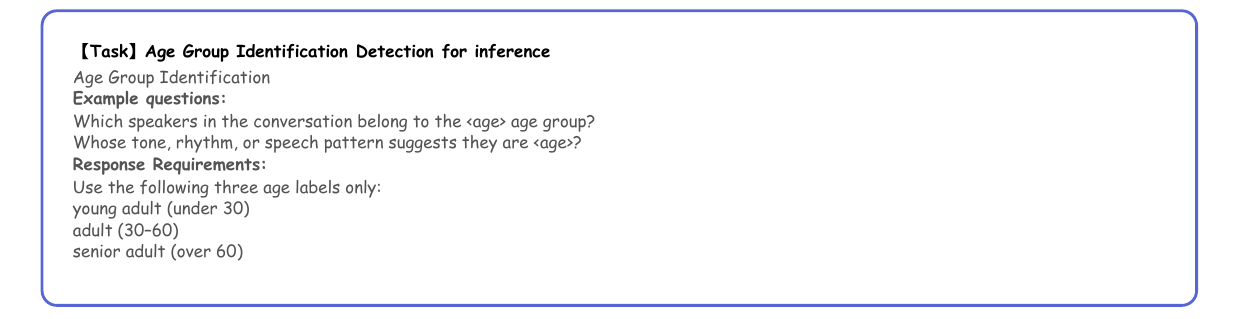}
\caption{Example Question and Requirements for Age Group Identification}
\label{fig:inference_age_group_identification}
\end{figure*}

\begin{figure*}[!htb]
\includegraphics[width=1.0\textwidth]{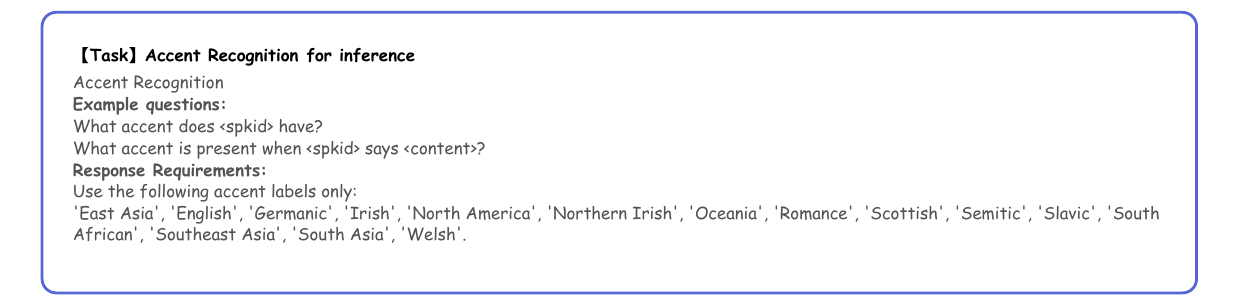}
\caption{Example Question and Requirements for Accent Recognition}
\label{fig:inference_accent_recognition}
\end{figure*}

\begin{figure*}[!htb]
\includegraphics[width=1.0\textwidth]{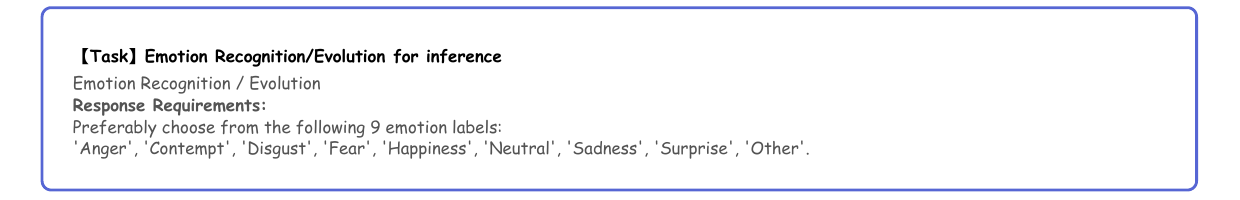}
\caption{Example Question and Requirements for Emotion Recognition/Evolution}
\label{fig:inference_emotion_recognition_evolution}
\end{figure*}

\begin{figure*}[!htb]
\includegraphics[width=1.0\textwidth]{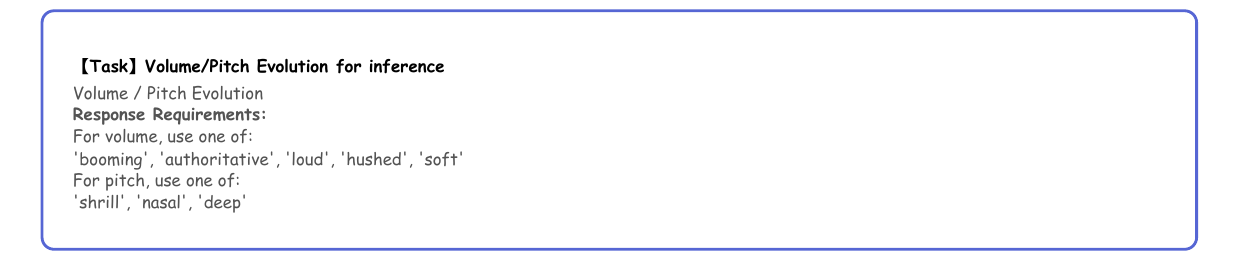}
\caption{Example Question and Requirements for Volume/Pitch Evolution}
\label{fig:inference_volume_pitch_evolution}
\end{figure*}

\begin{figure*}[!htb]
\includegraphics[width=1.0\textwidth]{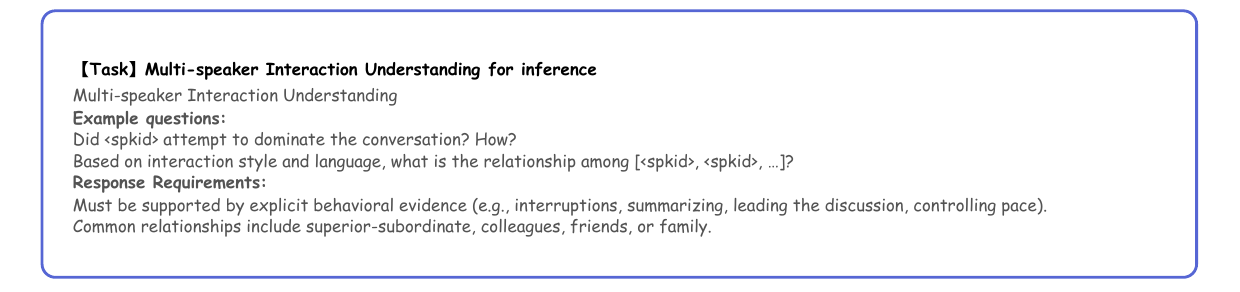}
\caption{Example Question and Requirements for Multi-Speaker Interaction Understanding}
\label{fig:inference_multi_speaker_interaction_understanding}
\end{figure*}

\begin{figure*}[!htb]
\includegraphics[width=1.0\textwidth]{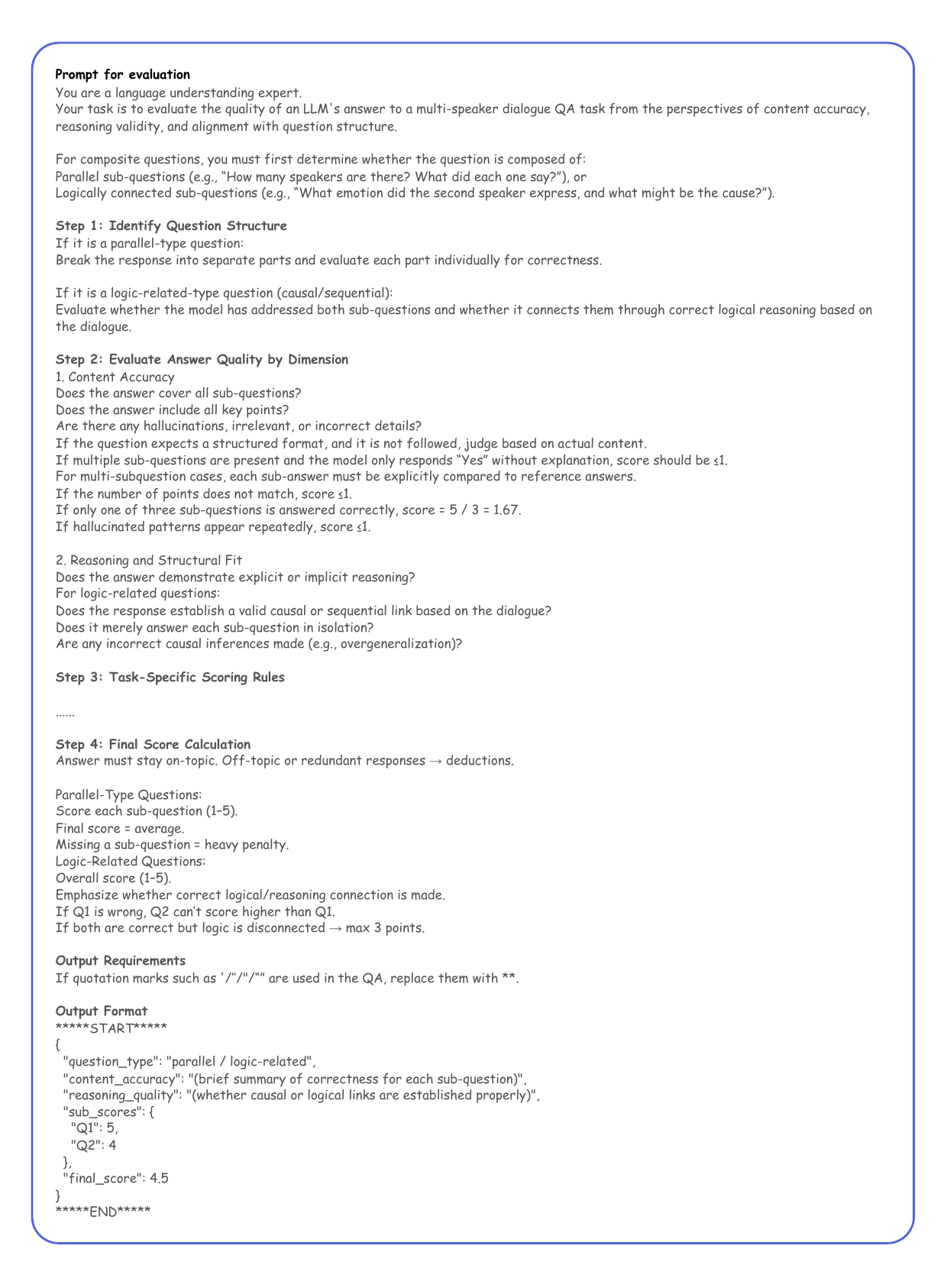}
\caption{LALM Evaluation Prompt}
\label{fig:evaluation_frame}
\end{figure*}

\begin{figure*}[!htb]
\includegraphics[width=1.0\textwidth]{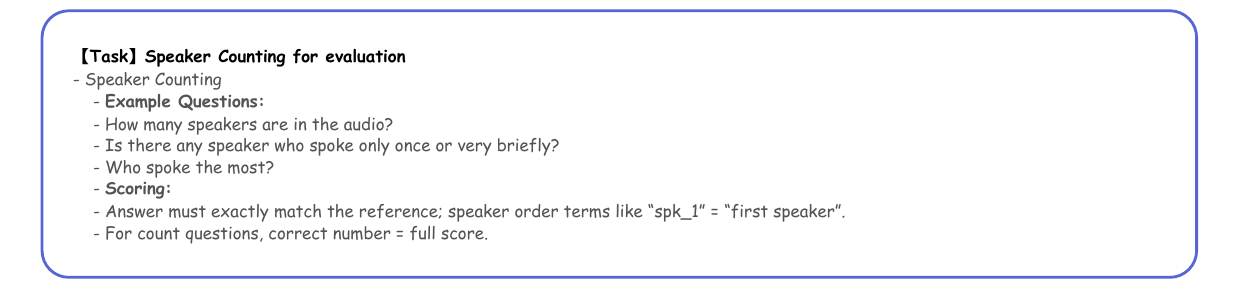}
\caption{Evaluation Requirements for Speaker Counting}
\label{fig:evaluation_speaker_counting}
\end{figure*}

\begin{figure*}[!htb]
\includegraphics[width=1.0\textwidth]{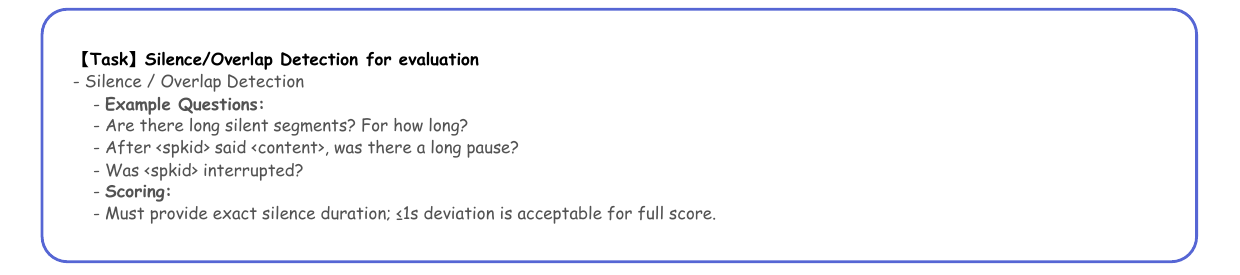}
\caption{Evaluation Requirements for Silence/Overlap Detection}
\label{fig:evaluation_silence_overlap_detection}
\end{figure*}

\begin{figure*}[!htb]
\includegraphics[width=1.0\textwidth]{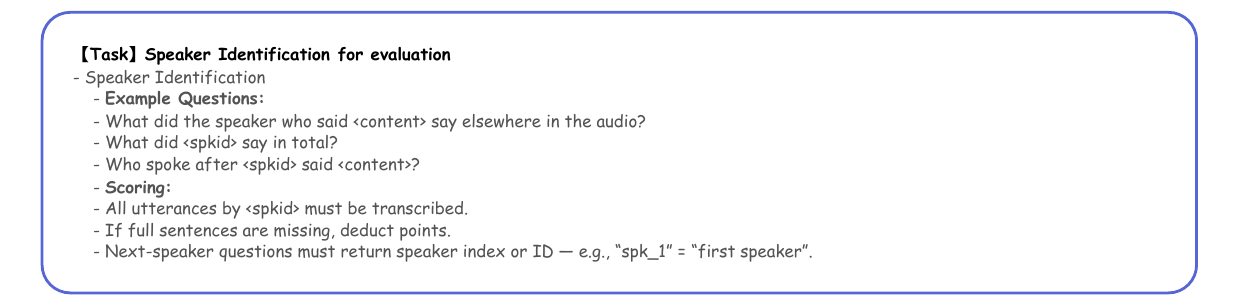}
\caption{Evaluation Requirements for Speaker Identification}
\label{fig:evaluation_speaker_identification}
\end{figure*}

\begin{figure*}[!htb]
\includegraphics[width=1.0\textwidth]{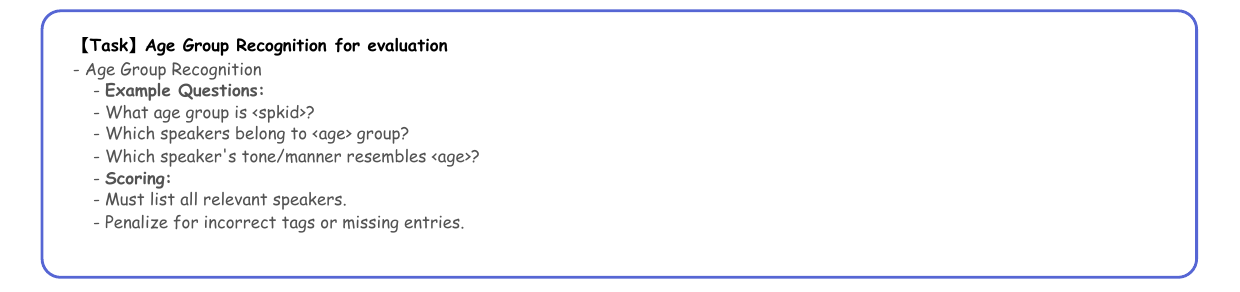}
\caption{Evaluation Requirements for Age Group Recognition}
\label{fig:evaluation_age_group_identification}
\end{figure*}

\begin{figure*}[!htb]
\includegraphics[width=1.0\textwidth]{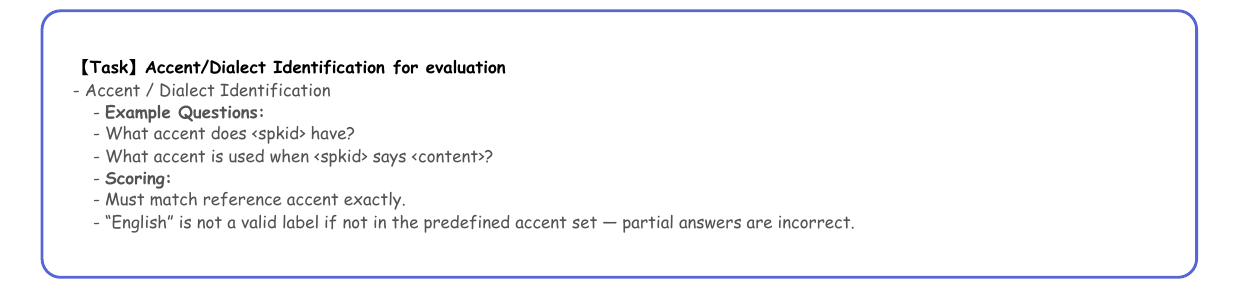}
\caption{Evaluation Requirements for Accent/Dialect Identification}
\label{fig:evaluation_accent_dialect_identification}
\end{figure*}

\begin{figure*}[!htb]
\includegraphics[width=1.0\textwidth]{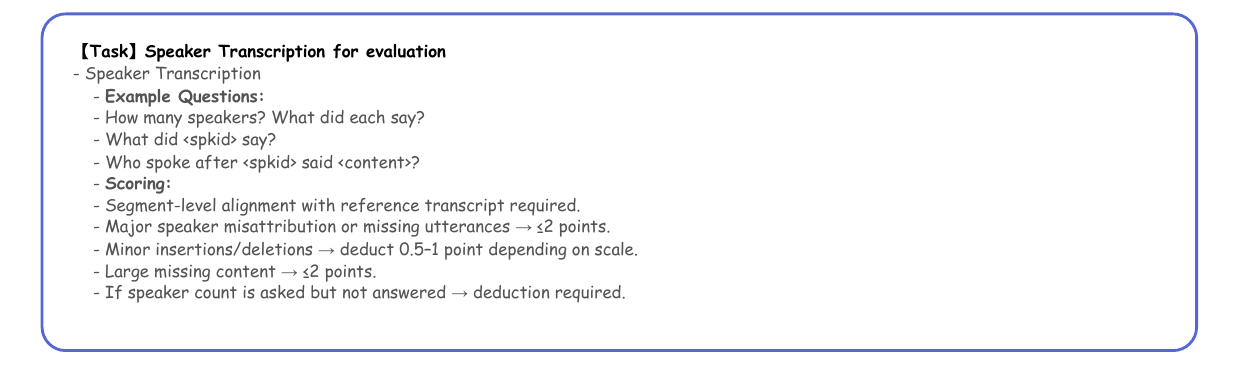}
\caption{Evaluation Requirements for Speaker Transcription}
\label{fig:evaluation_speaker_transcription}
\end{figure*}

\begin{figure*}[!htb]
\includegraphics[width=1.0\textwidth]{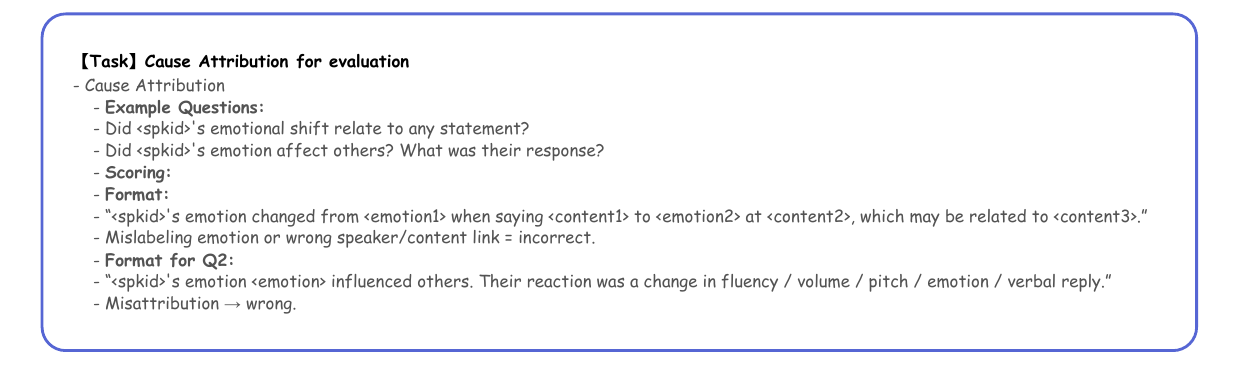}
\caption{Evaluation Requirements for Cause Attribution}
\label{fig:evaluation_cause_attribution}
\end{figure*}

\begin{figure*}[!htb]
\includegraphics[width=1.0\textwidth]{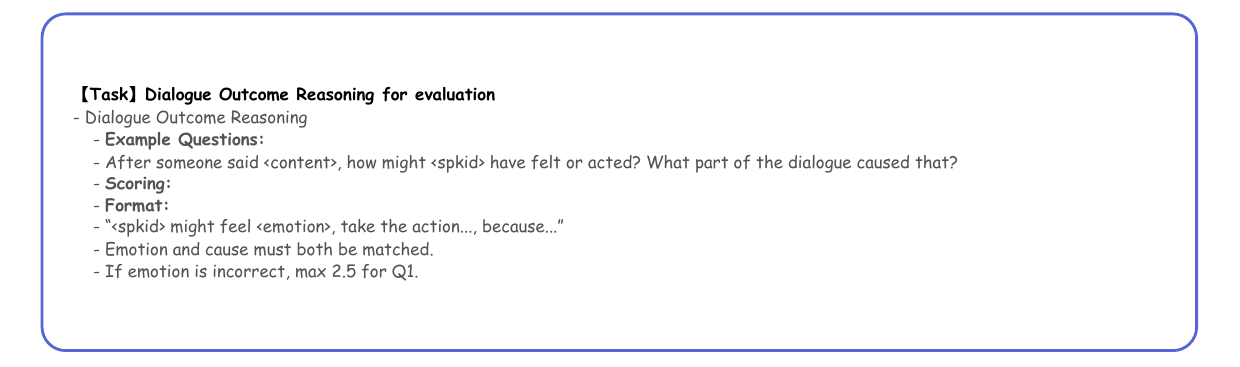}
\caption{Evaluation Requirements for Dialogue Outcome Reasoning}
\label{fig:evaluation_dialogue_outcome_reasoning}
\end{figure*}

% Detailed Performance Analysis of Different Models
While Figure 3 in the main text presents the performance of different models across various tasks in a unified visualization to facilitate cross-model comparison, we provide separate visualizations for each model's performance across different tasks in Figure~\ref{fig:ind_model} to enable clearer assessment of individual model capabilities

\centering
\onecolumn
\renewcommand{\arraystretch}{1.18}

\begin{longtable}[!t]{@{}p{0.14\textwidth} p{0.16\textwidth} p{0.24\textwidth} p{0.22\textwidth} p{0.14\textwidth}@{}}
\toprule
Error Type & Definition & Example & Prediction & Reason \\
\midrule
\endfirsthead

\toprule
Error Type & Definition & Example & Prediction & Reason \\
\midrule
\endhead

\addlinespace
\caption{Error type definitions with representative examples showing question contexts, expected answers (bold), model predictions, and failure analysis.}
\label{tab:error_definition} \\
\endlastfoot

Perceptual Errors & 
The model fails to perceive the audio accurately, resulting in an inaccurate or incomplete understanding of the input. & 
\begin{tabular}[t]{@{}p{0.24\textwidth}@{}}
How many female speakers are there in the audio? \\
\textbf{There is one female speaker in the audio.} 
\end{tabular} & Two & Model misperceived the gender characteristics of multiple speakers \\
\midrule
Reasoning Errors & 
The model understands the audio content but exhibits flaws in logical reasoning, leading to incorrect or unsound conclusions based on the input. & 
\begin{tabular}[t]{@{}p{0.24\textwidth}@{}}
What age group do the first speaker and second speaker most likely belong to? What commonly used expressions/topics of interest for that age group did they employ? \\
\textbf{Both the first speaker and second speaker are adults. Their discussion topics include English language learning, expression habits of non-native speakers, and how to conduct English conversations more naturally. These topics reflect adults' focus on language acquisition and cross-cultural communication, aligning with adult interests and thought patterns.} 
\end{tabular} & The first two speakers are both adults. They are discussing how to improve conversation skills. & Model fails to connect the age group with the question in its reasoning. \\
\midrule
Lack of Knowledge & 
The model understands the audio content but lacks the necessary knowledge or context to provide a correct or relevant answer. & 
\begin{tabular}[t]{@{}p{0.24\textwidth}@{}}
What are the accents of the first speaker and the second speaker, and how is this related to their viewpoints in the conversation? \\
\textbf{The accents of both the first speaker and the second speaker are East Asian accents. While discussing the issue of phone heating, the second speaker mentions 'middle frame heat is no big deal, what worries me is feeling concentrated heat on the upper part of the back cover after using it for a while'. This relates to the high level of attention consumers in East Asia pay to the heat dissipation performance of electronic products, reflecting the sensitivity of users in this region to product details and their emphasis on safety.} 
\end{tabular} & Based on your request, the analysis of the audio content is as follows: Both the first and second speakers have East Asian accents. The accent itself does not have a direct causal relationship with their viewpoints in the conversation. Their shared accent indicates that they belong to the same linguistic and cultural background, which provides a common context for their discussion. The differences in their viewpoints primarily stem from their individual concerns as consumers. Their accents merely place them within the same discussion circle, while their differing viewpoints arise from their distinct consumer positions and focal points. & Model lacks knowledge of consumer concerns about electronic products in the East Asian region. \\
\midrule
Rejection of Answer & 
The model fails to generate a response or refuses to provide an answer. & 
\begin{tabular}[t]{@{}p{0.24\textwidth}@{}}
Does the conversation ultimately move toward consensus or disagreement? Where can this be observed? \\
\textbf{Toward consensus. This is evident when the second speaker expresses *right, right, exactly, that's correct*, showing affirmation from others. Subsequently, both parties share similar life attitudes and viewpoints in the dialogue.} 
\end{tabular} & Empty string & Model fails to answer the question. \\
\midrule
Answer Extraction Errors & 
The model fails to follow instructions for answer extraction, resulting in an incorrectly formatted response. & 
\begin{tabular}[t]{@{}p{0.24\textwidth}@{}}
What was the first speaker's emotion when expressing *Hmm, before I thought, I didn't think there was any problem with this place*? Did it affect others? What was the second speaker's reaction? \\
\textbf{The first speaker's emotion when expressing *Hmm, before I thought, I didn't think there was any problem with this place* was Sadness, which affected others. The second speaker's reaction was emotional change/verbal response.} 
\end{tabular} & Sadness & Model only answers part of the question. \\
\bottomrule
\end{longtable}
\twocolumn

\end{document}